\documentclass[transmag]{IEEEtran}
\usepackage{latexsym}
\usepackage{graphicx}
\usepackage{amsfonts,amssymb,amsmath}
\usepackage{hyperref}
\usepackage{algorithmic, algorithm}
\usepackage{bm, dsfont, braket, amsmath, amssymb, textcomp, mathtools}
\usepackage{cleveref, cite}
\usepackage{subfig}
\usepackage{soul, color}

\usepackage{hhline}

\usepackage{dblfloatfix}

\DeclareMathOperator*{\argmin}{arg\,min}

\newcommand{\threeH}{0.3}

\def\BibTeX{{\rm B\kern-.05em{\sc i\kern-.025em b}\kern-.08em T\kern-.1667em\lower.7ex\hbox{E}\kern-.125emX}}
\begin{document}

\title{Approximate Message Passing with a Colored Aliasing Model for Variable Density Fourier Sampled Images}

\author{Charles Millard, \IEEEmembership{Student Member, IEEE}, Aaron T Hess, Boris Mailh\'{e} and Jared Tanner, \IEEEmembership{Member, IEEE}
\thanks{Manuscript submitted for review on 13th March 2020. A preliminary form of this manuscript was submitted to the IEEE International Conference on Image Processing 2020: see \cite{Millard2019}. This work was supported by an EPSRC Industrial CASE studentship with Siemens Healthineers, voucher number 17000051, and by The Alan Turing Institute under
the EPSRC grant EP/N510129/1. The concepts and information presented in this paper are based on research results that are not commercially available.}
\thanks{Charles Millard and Jared Tanner are with the Mathematical Institute at the University of Oxford, Oxford,  OX2 6GG, UK (e-mail: millard@maths.ox.ac.uk; tanner@maths.ox.ac.uk)}
\thanks{Aaron T Hess is with the Oxford Centre for Clinical Magnetic Resonance at the University of Oxford, Oxford,  OX3 9DU, UK (e-mail: aaron.hess@cardiov.ox.ac.uk)}
\thanks{Boris Mailh\'{e} is with Siemens Healthineers, 755 College Rd E, Princeton, NJ 08540, USA (e-mail: boris.mailhe@siemens-healthineers.com)}}

\IEEEtitleabstractindextext{\begin{abstract}
  The Approximate Message Passing (AMP) algorithm efficiently reconstructs signals which have been sampled with large i.i.d. sub-Gaussian sensing matrices. Central to AMP is its ``state evolution", which guarantees that the difference between the current estimate and ground truth (the ``aliasing") at every iteration obeys a Gaussian distribution that can be fully characterized by a scalar. However, when Fourier coefficients of a signal with non-uniform spectral density are sampled, such as in Magnetic Resonance Imaging (MRI), the aliasing is intrinsically colored, AMP's scalar state evolution is no longer accurate and the algorithm encounters convergence problems. In response, we propose the Variable Density Approximate Message Passing (VDAMP) algorithm, which uses the wavelet domain to model the colored aliasing. We present empirical evidence that VDAMP obeys a ``colored state evolution", where the aliasing  obeys a Gaussian distribution that can be fully characterized with one scalar per wavelet subband. A benefit of state evolution is that Stein's Unbiased Risk Estimate (SURE) can be effectively implemented, yielding an algorithm with subband-dependent thresholding that has no free parameters. We empirically evaluate the effectiveness of VDAMP on three variations of Fast Iterative Shrinkage-Thresholding (FISTA) and find that it converges in around 10 times fewer iterations on average than the next-fastest method, and to a comparable mean-squared-error.    
\end{abstract}

}

\maketitle

\section{Introduction}

\IEEEPARstart{W}e consider a complex data vector $\bm{y} \in \mathds{C}^{N}$ formed of noisy Fourier coefficients of a deterministic signal of interest $\bm{x}_0 \in \mathds{C}^N$:
\begin{equation}
\bm{y} = \bm{M}_\Omega (\bm{F x}_0 + \bm{\varepsilon}), \label{eqn:MRI_regression} 
\end{equation}
where $\bm{F}$ is a multi-dimensional discrete Fourier transform and $\bm{M}_\Omega \in \mathds{R}^{N \times N}$ is a diagonal undersampling mask with $1$ on the $j$th diagonal entry if $j \in \Omega$ and $0$ otherwise, where $\Omega$ is a sampling set with $|\Omega| = n$ for $n <N$. Here, $\bm{\varepsilon} \backsim \mathcal{CN}(\bm{0}, \sigma_\varepsilon^2 \mathds{1}_N)$ where $\mathds{1}_N$ is the $N \times N$ identity matrix and $\mathcal{CN}(\bm{\mu}, \bm{\Sigma}^2)$ denotes the distribution with independent  real and imaginary parts that are normally distributed with mean $\bm{\mu}$ and covariance matrix $\bm{\Sigma}^2/2$. A well-studied approach is to seek a solution of 
\begin{equation}
   \hat{\bm{x}} = \underset{\bm{x} \in \mathds{C}^N}{\operatorname{argmin}} \frac{1}{2} \| \bm{y} -  \bm{M}_\Omega \bm{F}\bm{x}\|^2_2 + f(\bm{x}) \label{eqn:recon}
\end{equation}
where $f(\bm{x})$ is a model-based penalty function. Compressed sensing \cite{Donoho2006, Candes2006} concerns the reconstruction of signals of interest from underdetermined measurements, where sparsity in $\hat{\bm{x}}$ is promoted by solving \eqref{eqn:recon} with $f(\bm{x}) = \lambda \|\bm{\Psi x}\|_1$ for sparse weighting $\lambda > 0$ and sparsifying transform $\bm{\Psi}$. 

A prominent success of compressed sensing with Fourier measurements is accelerated Magnetic Resonance Imaging (MRI) \cite{Lustig2007, Otazo2010, Jaspan2015, Ye2019, Donoho2017}. Images of interest typically have a highly non-uniform spectral density that is concentrated at low frequencies. Accordingly, it is well-known that better image restoration is possible if the sampling set $\Omega$ is generated with variable density, so that there is a higher probability of sampling low frequencies  \cite{Puy2011, wangVDS, Krahmer2014, Chauffert2013, Adcock2017}. This work considers an $\Omega$ with elements drawn independently from a Bernoulli distribution with generic non-uniform probability, so that $\mathrm{Prob}(j \in \Omega) = p_{j} \in [0,1]$.

\subsection{Approximate Message Passing}
 The Approximate Message Passing (AMP) algorithm  \cite{Donoho2009} is an iterative method that, for certain sensing matrices $\bm{\Phi} \in \mathds{R}^{n \times N}$, efficiently estimates $\bm{x}_0$ in problems of the form
$\bm{y} = \bm{\Phi}\bm{x}_0 + \bm{\varepsilon}. 
$
At iteration $k$, AMP implements a denoiser $\bm{g}(\bm{r}_k; \tau_k)$ on $\bm{r}_{k}$ with mean-squared error estimate $\tau_k$, which can be, for instance, the proximal operator associated with penalty function $f(\bm{x})$:
\begin{equation}
    \bm{g}(\bm{r}_k; \tau_k) = \underset{\bm{x} \in \mathds{C}^N}{\operatorname{argmin}} \frac{1}{2\tau_k}\|\bm{r}_k - \bm{x}\|^2_2 + f(\bm{x}), \label{eqn:prox_denoi}
\end{equation}
which is equal to soft thresholding in the case of $f(\bm{x}) = \lambda \|\bm{\Psi x}\|_1$ and orthogonal $\bm{\Psi}$. Under certain circumstances \cite{Donoho2009}, AMP with proximal denoising shares a fixed point with optimization problems of the form of \eqref{eqn:recon}. Further, for certain sensing matrices and given mild conditions on $f(\bm{x})$, AMP obeys a \textit{state evolution}, which guarantees that in the large system limit $n,N \rightarrow \infty$, $n/N \rightarrow \delta \in (0,1)$, vector $\bm{r}_{k}$  is the original signal corrupted by zero-mean Gaussian noise with a covariance matrix that is proportional to the identity:  
\begin{equation}
    \bm{r}_{k} = \bm{x}_0 + \mathcal{N}(\bm{0}, \sigma_k^2\mathds{1}_N), \label{eqn:SE}
\end{equation}
where  $\sigma_k$ is a scalar iteration-dependent standard deviation. In this work, the term \textit{aliasing} is used to refer to the difference between a given estimate and the ground truth. For instance, the aliasing of $\bm{r}_k$ is $\bm{r}_k - \bm{x}_0$. Also, when the covariance matrix is proportional to the identity, as in \eqref{eqn:SE}, the aliasing and state evolution are referred to as \textit{white}. 

AMP was originally constructed for real, zero-mean, i.i.d. Gaussian measurements, and its white state evolution  was proven for this case in \cite{Bayati2011} and subsequently proven for i.i.d. sub-Gaussian measurements in \cite{Bayati2015}. It has also been  shown empirically that it holds for uniformly undersampled Fourier measurements of an artificial i.i.d. signal \cite{Donoho2009}. When state evolution holds, AMP is known to exhibit very fast convergence. However, for generic $\bm{\Phi}$, the behavior of AMP is not well understood and it has been noted by a number of authors \cite{Rangan2014, Caltagirone2014,Guo2015, Rangan2016} that it can encounter convergence problems. The recent Orthogonal AMP (OAMP) \cite{Ma2017}  and related Vector Approximate Message Passing (VAMP) \cite{Rangan2019} algorithm obey a white state evolution for a broader class of measurement matrices $\bm{\Phi}$, and were found to perform very well on certain reconstruction tasks. For VAMP, white state evolution was proven for sensing matrices that are ``right-orthogonally invariant": see \cite{Rangan2019} for details.

\begin{algorithm}[t]
\caption{OAMP  \cite{Ma2017} \label{alg:OAMP}}
\textbf{Require:} Matrix $\bm{\Phi}$, measurements $\bm{y}$, denoiser $\bm{g}(\bm{r}_{k}; \tau_{k})$, number of iterations $K_{it}$.
\begin{algorithmic}[1]
\STATE Set $\widetilde{\bm{r}}_{0} = \bm{0}$ 
\FOR {$k =0,1,\ldots, K_{it}-1$} 
\STATE $\bm{z}_k = \bm{y} -  \bm{\Phi}\widetilde{\bm{r}}_{k}$ 
\STATE $\bm{r}_{k} = \widetilde{\bm{r}}_{k} + \bm{\Phi}^H \bm{z}_k$ 
\STATE Update $\tau_k$ 
\STATE $\hat{\bm{x}}_k = \bm{g}(\bm{r}_{k}; \tau_{k})$ 
\STATE $\alpha_k = \braket{\bm{g}'(\bm{r}_{k}; \tau_{k})}$
\STATE Update $c_k$, e.g. $c_k = 1, 2 \text{ or } 3$
\STATE $\widetilde{\bm{r}}_{k+1} = c_k \cdot (\hat{\bm{x}}_k - \alpha_k \bm{r}_{k})$ 
\ENDFOR
\RETURN $\hat{\bm{x}}_k$ 
\end{algorithmic}
\end{algorithm}

The matched filter variation of the OAMP algorithm \cite{Ma2017}, which forms the basis of the algorithm presented in this work, is stated in Algorithm \ref{alg:OAMP}. Here, $\braket{\cdot}$ is the empirical averaging operator and $\bm{g}'(\bm{r}_{k}; \tau_{k})$ is the diagonal of the Jacobian of $\bm{g}(\bm{r}_{k}; \tau_{k})$ with respect to $\bm{r}_k$. The scalar $\tau_k$ estimates the variance $\sigma_k^2$ from \eqref{eqn:SE}: see Eqn. (31) of \cite{Ma2017} for details of the update formula.  The relationship between OAMP and the well-known Iterative Shrinkage-Thresholding Algorithm (ISTA) \cite{Daubechies2004} can be seen by considering lines 6-9 as a single function:
\begin{equation}
    \tilde{\bm{g}}(\bm{r}_{k}; \tau_{k}) = c_k \cdot (\bm{g}(\bm{r}_{k}; \tau_{k}) - \alpha_k \bm{r}_{k}). \label{eqn:g_tilde_oamp}
\end{equation} 
Then OAMP is equivalent to ISTA with $\tilde{\bm{g}}(\bm{r}_{k}; \tau_{k})$ in place of the usual shrinkage step. The $ \alpha_k \bm{r}_{k} $ subtraction, known as the \textit{Onsager correction} \cite{Donoho2009},  causes the function $\tilde{\bm{g}}(\bm{r}_{k}; \tau_{k})$ for large $N$ to be approximately \textit{divergence-free}  \cite{Ma2017}, defined as
\begin{equation}
	\braket{\tilde{\bm{g}}'(\bm{r}_{k}; \tau_{k})} \approx 0. \label{eqn:div_free}
\end{equation}
The divergence-free property of $\tilde{\bm{g}}(\bm{r}_{k}; \tau_{k})$  is the vital aspect of OAMP that leads to the white state evolution of  \eqref{eqn:SE} \cite{Ma2017}. Although any divergence-free function can be employed in place of lines 6-9 of Algorithm \ref{alg:OAMP}, and is not required to take the form of \eqref{eqn:g_tilde_oamp}, this work focuses on a $\tilde{\bm{g}}(\bm{r}_{k}; \tau_{k})$ of the form of \eqref{eqn:g_tilde_oamp} with $\bm{g}(\bm{r}_{k}; \tau_{k})$ as the soft thresholding operator. 

Any choice of $c_k$ update in line 8 of Algorithm \ref{alg:OAMP} is consistent with the divergence-free requirement of \eqref{eqn:div_free}. For soft thresholding, OAMP \cite{Ma2017} gives no explicit suggestions for $c_k$ in practice. Instead, OAMP demonstrates the generality of its state evolution using three arbitrary chosen values $c_k = 1, 2, 3$. For the algorithm presented in this work, two $c_k$ updates are suggested, stated in \eqref{eqn:Cvamp} and \eqref{eqn:Csure}, which are motivated by VAMP \cite{Rangan2019} and computed by Stein's Unbiased Risk Estimate (SURE) \cite{Stein1981, Xue2016} respectively.

\subsection{Colored aliasing\label{sec:colored_aliasing}}
AMP, OAMP and VAMP assume that the sensing matrix is sufficiently random to ensure that the aliasing is white. However, when an image is sampled in the Fourier domain, the aliasing is innately colored. To see this, consider the natural initialization for an approximate message passing algorithm: the unbiased estimator $\widetilde{\bm{x}} = \bm{F}^{H}\bm{P}^{-1}\bm{y}$, where $\bm{P}$ is the  diagonal  matrix  formed  from  sampling  probabilities $p_j$. Denoting $\bm{y}_0 = \bm{F x}_0$, the power spectrum of the aliasing of $\widetilde{\bm{x}}$ is shown in Appendix \ref{app:power_spec} to be
\begin{align}
    \mathds{E}_{\Omega, \varepsilon} \{ |\bm{y}_0 - \bm{P}^{-1}\bm{y}|^2\} & = ( \bm{P}^{-1} - \mathds{1}_N)|\bm{y}_0|^2 + \sigma^2_{\varepsilon}   \bm{P}^{-1} \bm{1}_N \label{eqn:colored_al}
\end{align}
where  $|\cdot|$ is the entry-wise absolute value and $\bm{1}_N$ is the $N$-dimensional vector of ones. Equation \eqref{eqn:colored_al} depends on $\bm{P}$ and $|\bm{y}_0|^2$, which are non-uniform and anisotropic in general. Note that although the specific case $p_j = (\sigma^2_{\varepsilon}  + |y_{0,j}|^2)/(\alpha + |y_{0,j}|^2)$ for constant $\alpha$ does lead to white aliasing, it requires knowledge of the ground truth spectral density $|\bm{y}_0|^2$ so is not a feasible sampling scheme in practice.  

\begin{figure}[t!]
\captionsetup[subfigure]{labelformat=empty}
\centering

   \subfloat[\label{fig:shepp_logan}(a) $\bm{x}_0$]{%
      \includegraphics[height = 0.27 \columnwidth]{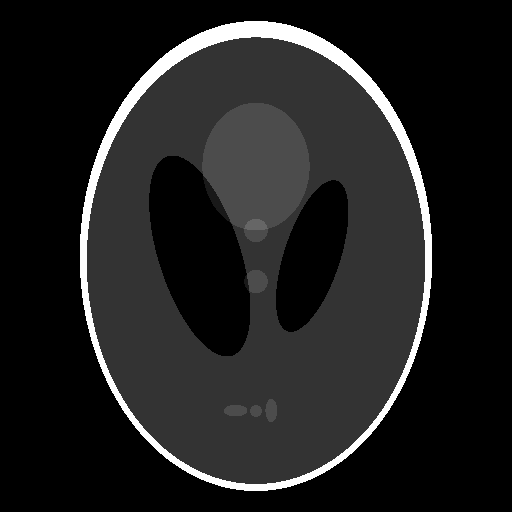}}
      \hspace{0.1cm}
   \subfloat[\label{fig:x_hat}(b) $\widetilde{\bm{x}}$]{%
      \includegraphics[height=0.27 \columnwidth]{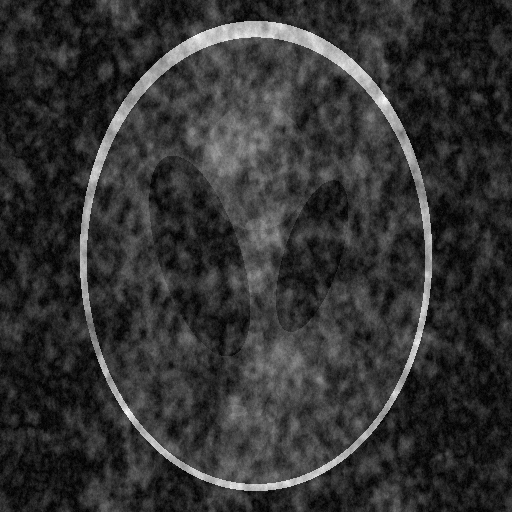}}
    \hspace{0.1cm}
   \subfloat[\label{fig:x_res}(c) $|\bm{x}_0 - \widetilde{\bm{x}}|$]{%
      \includegraphics[height = 0.27 \columnwidth]{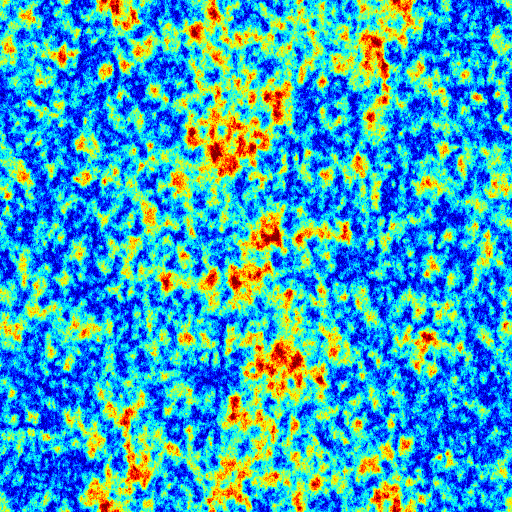}}
    \hspace{0.1cm}
   \subfloat[]{%
      \includegraphics[height=0.26 \columnwidth]{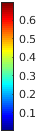} \captionsetup[subfigure]{labelformat=empty}  \nonumber}
      \hspace{\fill} \\
    \hspace{0.02cm}
   \subfloat[\label{fig:w0}(d) $\bm{w}_0$]{%
      \includegraphics[height = 0.27 \columnwidth]{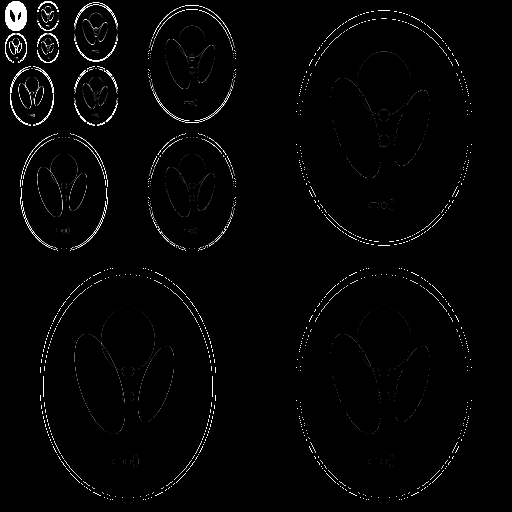}}
      \hspace{0.1cm}
   \subfloat[\label{fig:w_hat}(e) $\bm{r}_0$]{%
      \includegraphics[height=0.27 \columnwidth]{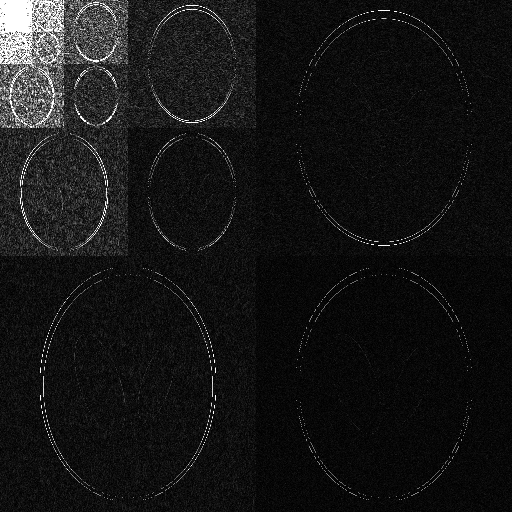}}
      \hspace{0.1cm}
   \subfloat[\label{fig:w_res}(f) $|\bm{r}_0 - \bm{w}_0|$]{%
      \includegraphics[height=0.27 \columnwidth]{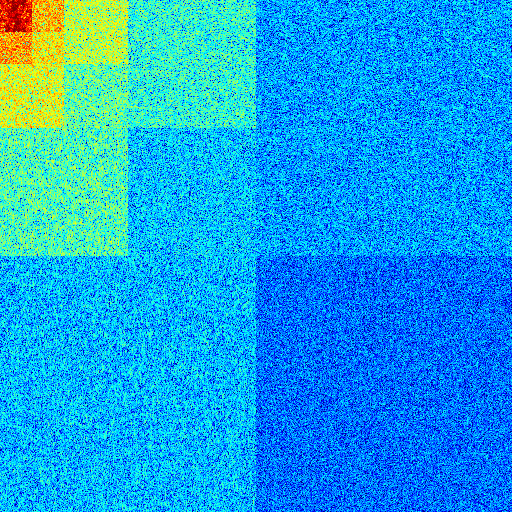}}
      \hspace{0.1cm}
   \subfloat[]{%
      \includegraphics[height=0.27  \columnwidth]{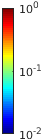} \captionsetup[subfigure]{labelformat=empty} }
      \hspace{\fill} 
      \caption{\label{fig:intuition} The ground truth, unbiased estimate and entry-wise absolute error for a uniformly sampled Shepp-Logan with $p_j = 1/2$, where the colorbars are given as a proportion of the maximum of $\bm{x}_0$. The top row shows the image domain and the bottom shows the wavelet domain. The colored aliasing evident in Fig. \ref{fig:x_res} illustrates the infeasibility of white state evolution for Fourier sampling of signals with non-uniform spectral density. The anisotropy of the spectral density of $\bm{x}_0$ causes the horizontal, vertical and diagonal variances to differ, even at the same scale.}
\end{figure}

A visual example of colored aliasing is shown in the top row of Fig. \ref{fig:intuition}. Here, the unbiased estimate of a 512x512 synthetic Shepp-Logan uniformly sampled with $p_j = 1/2$ for all $j$ is shown, with $\sigma_{\varepsilon} = 0$. By \eqref{eqn:colored_al}, the power spectrum of the aliasing of $\tilde{\bm{x}}$ in this case is $\mathds{E}_{\Omega, \varepsilon} \{ |\bm{y}_0 - \bm{P}^{-1}\bm{y}|^2\} = |\bm{y}_0|^2$. Since $|\bm{y}_0|^2$ is tightly concentrated at low frequencies, the aliasing has strong low-frequency components, which is manifest in Fig. \ref{fig:x_res} as local correlations. In this example, uniform sampling was chosen to exaggerate the colored aliasing property. In practice, when variable density sampling is used, the sampling distribution partially compensates for the power spectrum of the signal, so the aliasing is still colored, but less strongly. 

The intrinsically colored aliasing of variable density sampling from a non-uniform spectral density implies that the white state evolution of AMP, OAMP, and VAMP, \eqref{eqn:SE}, cannot be relied upon. The primary development of this work is based on the use of the Discrete Wavelet Transform (DWT) to compute a multiresolution decomposition of the power spectrum of the aliasing, as used for colored noise analysis in \cite{Goossens2011, Li2010, Johnstone1997}. In the wavelet domain, colored aliasing has a structure that resembles a state evolution. To illustrate this, consider again the unbiased initialization $\widetilde{\bm{x}}$. The corresponding estimator in the wavelet domain is
\begin{equation}
\bm{r}_0 = \bm{\Psi} \widetilde{\bm{x}} \label{eqn:r0}.
\end{equation}
The bottom row of Fig. \ref{fig:intuition} shows $\bm{w}_0 =\bm{\Psi x}_0$, $\bm{r}_0$ and the entry-wise absolute difference $|\bm{w}_0 -\bm{r}_0|$ for the same image and sampling set as the top row, where $\bm{\Psi}$ is a Haar DWT with 4 decomposition scales. Qualitatively, Fig. \ref{fig:intuition}f suggests that the power spectrum given by \eqref{eqn:colored_al} has a simple structure in the wavelet domain; in particular, it suggests that the per-subband aliasing within each subband is approximately uniform.  In Section \ref{sec:NumExp} we show that, in fact, \textit{the aliasing within each subband is quantitatively consistent with a white Gaussian distribution}, see Fig. \ref{fig:qqplots}. Further, Table \ref{tab:mean_kurt_re} presents evidence that the Gaussianity holds for a variety of image types and undersampling factors. 

\subsection{Colored state evolution}

 Herein we present a new method for undersampled signal reconstruction that we term the Variable Density Approximate Message Passing (VDAMP) algorithm, see Algorithm \ref{alg:VDAMP}.   We present empirical evidence that VDAMP preserves the subband-dependent noise structure illustrated in Fig. \ref{fig:intuition}f for all iterations. Explicitly, the $\bm{r}_{k}$ of VDAMP behaves as 
\begin{equation}
 \bm{r}_{k} = \bm{w}_0 + \mathcal{CN}(\bm{0},  \bm{\Sigma}_k^2 ), \label{eqn:VDAMPse}
\end{equation}
where $\bm{\Psi}$ is an orthogonal DWT and the covariance matrix $\bm{\Sigma}_k^2$ is diagonal so that for a $\bm{\Psi}$ with $s$ decomposition scales,
\begin{equation}
    \bm{\Sigma}_k^2 = \begin{bmatrix}
    \sigma^2_{k, 1} \mathds{1}_{N_{1}} & 0 & \dots & 0 \\
    0       &  \sigma^2_{k, 2} \mathds{1}_{N_2}& \dots & 0 \\
    \vdots       &  \vdots & \ddots &  \vdots \\
    0       & 0 & \hdots &  \sigma^2_{k, 1+3s}\mathds{1}_{N_{1+3s}}
    \end{bmatrix}, \label{eqn:sigma}
\end{equation}
where $\sigma^2_{k, b}$ and $N_b$ refer to the variance and dimension of the $b$th subband respectively. We refer to \eqref{eqn:VDAMPse} and \eqref{eqn:sigma} as the \textit{colored state evolution} of VDAMP and the aliasing of $\bm{r}_k$ as the \textit{effective noise} of VDAMP. 

The joint space-frequency localization provided by the wavelet transform decomposes the color of the effective noise while retaining incoherence. The algorithm presented in this work considers a sparse model on wavelet coefficients, however, we emphasize that the wavelet transform is primarily used as a tool for decomposing the aliasing, and is not necessarily constrained to model wavelet coefficients directly \cite{Metzler2016, Xue2016}. 

VAMP for Image Recovery (VAMPire) \cite{Schniter2017a} is an adaption of VAMP for variable density Fourier sampled images that uses wavelets to decompose the effective noise in to frequency bands that are subsequently ``whitened'' by a hand-tuned prediction of the per-subband energy. As in OAMP, Algorithm \ref{alg:OAMP}, the effective noise model of VAMPire is represented by a scalar $\tau_k$. In contrast, we propose making the necessary algorithmic adaptions to \textit{allow the aliasing to be colored}, and to model the color with a \textit{vector} $\bm{\tau}_k$.  To our knowledge, VDAMP is the first algorithm for variable density Fourier sampling of images where a state evolution has been observed.

\section{Description of algorithm\label{sec:VDAMP}}

\begin{algorithm}[t]
\caption{VDAMP \label{alg:VDAMP}}
\textbf{Require:} Sampling set $\Omega$, wavelet transform $\bm{\Psi}$, probability matrix $\bm{P}$, measurements $\bm{y}$, denoiser $\bm{g}(\bm{r}_{k}; \bm{\tau}_{k})$, number of iterations $K_{it}$.
\begin{algorithmic}[1]
\STATE Set $\widetilde{\bm{r}}_{0} = \bm{0}$ and compute $\bm{S} = |\bm{F}\bm{\Psi}^H|^2$
\FOR {$k =0,1,\ldots, K_{it}-1$} 
\STATE $\bm{z}_k = \bm{y} - \bm{M}_\Omega \bm{F} \bm{\Psi}^H\widetilde{\bm{r}}_{k}$ \label{algline:3}
\STATE $\bm{r}_{k} = \widetilde{\bm{r}}_{k} + \bm{\Psi} \bm{F}^H \bm{P}^{-1}\bm{z}_k$ \label{algline:4}
\STATE $\bm{\tau}_{k} = \bm{S}^H \bm{M}_\Omega \bm{P}^{-1} [(\bm{P}^{-1} - \mathds{1}_N)|\bm{z}_k|^2 + \sigma_\varepsilon^2 \bm{1}_N] $ \label{algline:5}
\STATE $\hat{\bm{w}}_{k} = \bm{g}(\bm{r}_{k}; \bm{\tau}_{k})$ \label{algline:6}
\STATE $\bm{\alpha}_{k} = \braket{\bm{\partial} (\bm{g}(\bm{r}_{k}; \bm{\tau}_{k}))}_\mathrm{sband}$ \label{algline:7}
\STATE Update $\bm{c}_k$ \label{algline:8}
\STATE $\widetilde{\bm{r}}_{k+1} = \bm{c}_k \odot (\hat{\bm{w}}_{k} - \bm{\alpha}_k \odot \bm{r}_{k})$ \label{algline:9}
\ENDFOR
\RETURN $\hat{\bm{x}} = \bm{\Psi}^H\hat{\bm{w}}_k + \bm{F}^H (\bm{y} - \bm{M}_\Omega\bm{F} \bm{\Psi}^H \hat{\bm{w}}_k)$ \label{algline:11}
\end{algorithmic}
\end{algorithm}

The VDAMP algorithm, Algorithm \ref{alg:VDAMP}, adapts OAMP, Algorithm \ref{alg:OAMP}, to a colored aliasing model. VDAMP's colored aliasing model is updated in line 5, where the scalar $\tau_k \in \mathds{R} $ of OAMP is replaced by a vector $\bm{\tau}_k \in \mathds{R}^N$ that models the diagonal of $\bm{\Sigma}_k^2$. In line 6, OAMP's denoiser $\bm{g}(\bm{r}_{k}; \tau_{k})$ is replaced by the denoiser $\bm{g}(\bm{r}_{k}; \bm{\tau}_{k})$, which takes the vector $\bm{\tau}_{k}$ as its input. Lines 7-9 of Algorithm \ref{alg:VDAMP} is the Onsager correction from lines 7-9 of Algorithm \ref{alg:OAMP} adapted to a subband-wise aliasing model. Here, the notation $\bm{\partial}(\bm{g}(\bm{r}_{k}; \bm{\tau}_{k}))$ in line 7 replaces OAMP's $\bm{g}'(\bm{r}_{k}; \tau_{k})$, defined as the function with $j$th entry
\begin{equation}
   \partial_j (\bm{g}(\bm{r}_{k}; \bm{\tau}_{k})) =  \frac{1}{2}\left(\frac{\partial  \Re[g_j(\bm{r}_k; \bm{\tau}_{k})]}{\partial \Re[r_j]}  + \frac{\partial \Im [g_j(\bm{r}_k; \bm{\tau}_{k})]}{\partial \Im [r_j]} \right) \label{eqn:partial}
\end{equation}
where $\Re [\cdot]$ and $\Im[\cdot]$ are the real and imaginary parts respectively. The form of \eqref{eqn:partial} is justified in section \ref{sec:VDAMP_ons_corr}. Also in line 7, the notation $\braket{\cdot}_{\mathrm{sband}}$ is an operator that empirically averages entries within subbands, so that $\bm{\alpha}_{k}$ has the structure 
\begin{equation}
 \bm{\alpha}_k = \begin{bmatrix}
    \alpha_{k,1} \bm{1}_{N_1} \\
    \alpha_{k,2} \bm{1}_{N_2}\\
    \vdots \\
    \alpha_{k,1+3s} \bm{1}_{N_{1+3s}}
    \end{bmatrix}  \label{eqn:vecalpha}
\end{equation}
with
\begin{equation}
    \alpha_{k,b} = \frac{1}{N_b} \sum_{j \in J_b}  \partial_j (\bm{g}(\bm{r}_{k}; \bm{\tau}_{k})),
\end{equation}
 where $J_b$ is the set of indices associated with subband $b$ and $N_b = |J_b|$. In line 8 of Algorithm \ref{alg:VDAMP}, $\bm{c}_k$ is a vector with the piecewise-constant structure of \eqref{eqn:vecalpha}, and in line 9 the notation $\odot$ refers to entry-wise multiplication. The remainder of this section works through Algorithm \ref{alg:VDAMP} in detail, line-by-line. 

\subsection{Density compensated gradient descent, lines 3-4}

To ensure that $\bm{r}_{k}$ is an unbiased estimate of $\bm{x}_0$, the sensing matrix must be correctly normalized. In VDAMP this is manifest in the gradient step of lines 3-4, which features a crucial weighting by $\bm{P}^{-1}$ that is absent in previous applications of AMP to variable density sampling \cite{Sung2013, Eksioglu2018,Schniter2017a}, where a state evolution was not observed. This provides the correct normalization in expectation over $\Omega$: $\mathds{E}_\Omega\{\bm{\Psi} \bm{F}^H \bm{P}^{-1} \bm{M}_\Omega \bm{F \Psi}^H\} = \mathds{1}_N$. Note that VDAMP's $\bm{r}_0$ is the unbiased estimator from \eqref{eqn:r0}. Such a rescaling is referred to as \textit{density compensation} in the MRI literature \cite{Pipe1999, Pruessmann2001}, and was used in the original compressed sensing MRI paper with zero-filling  to generate a unregularized, non-iterative baseline \cite{Lustig2007}. However, to our knowledge, VDAMP is the first \textit{iterative} method that employs density compensated gradient descent. Density compensation also arises in recovery guarantees for variable density Fourier measurements in \cite{Krahmer2014, Puy2011}, although it was considered an artifact of the proof and was not used in the numerical evaluations of these works.  The connection of VDAMP to these theoretical results is beyond the scope of this paper and is left as future work. 

Density compensation increases the variance of the measurement noise at frequencies sampled with low probability, and its inclusion in the gradient step will lead to a $\bm{r}_k$ with higher mean-squared error than an unweighted gradient step. However, as shown in Section \ref{sec:NumExp}, a careful choice of denoiser $\bm{g}(\bm{r}_{k}; \bm{\tau}_{k})$ that leverages VDAMP's state evolution can cause lines 6-9  to be very effective, leading to faster overall convergence than competing methods.   

The final step of VDAMP, line 11, is a gradient step without a $\bm{P}^{-1}$ weighting, which generates a biased image estimate $\hat{\bm{x}}$ with high data fidelity.

\subsection{Colored effective noise model, line 5}

Line 5 of Algorithm \ref{alg:VDAMP} computes an estimate of the colored effective noise covariance matrix $\bm{\Sigma}_k^2$ from \eqref{eqn:VDAMPse}. Through a similar derivation to that for \eqref{eqn:colored_al}, shown in Appendix \ref{app:power_spec}, the power spectrum of the aliasing of $\bm{r}_{k}$ is
\begin{multline}
    \mathds{E}_{\Omega,\varepsilon}\{| \bm{F}\bm{\Psi}^H \bm{r}_k - \bm{y}_0 |^2\} = (\bm{P}^{-1} - \mathds{1}_N) |\bm{F}\bm{\Psi}^H \widetilde{\bm{r}}_k - \bm{y}_0 |^2 \\ + \sigma_\varepsilon^2 \bm{P}^{-1}\bm{1}_N. \label{eqn:exp_y}
\end{multline}
Eqn. \eqref{eqn:exp_y} depends on the ground truth $\bm{y}_0$, so is of limited practical use. An estimate of \eqref{eqn:exp_y} that does not require knowledge of $\bm{y}_0$ is
\begin{equation}
     \bm{\tau}^y_k = \bm{M}_\Omega \bm{P}^{-1}[(\bm{P}^{-1} - \mathds{1}_N) |\bm{z}_k|^2 + \sigma_\varepsilon^2 \bm{1}_N]. \label{eqn:tauk}
\end{equation}
Estimating properties of a distribution using samples from another is known as \textit{importance sampling} in the statistics literature \cite{gamerman2006markov, chatterjee2018sample}.  Eqn. \eqref{eqn:tauk} uses importance sampling with $\bm{P}$ as the importance distribution which, as proven in Appendix \ref{app:unbiased_tau_update}, is an unbiased estimator of \eqref{eqn:exp_y}. We assume the estimator $\bm{\tau}_k^y$ concentrates around its expectation, and leave the study of how this depends on the importance distribution $\bm{P}$ for future works.

The computation of $\bm{\tau}_k$ in line 5 of Algorithm \ref{alg:VDAMP} is a linear transform of $\bm{\tau}^y_k$ to the wavelet domain: $\bm{\tau}_k = |\bm{\Psi F}^H|^2 \bm{\tau}^y_k$, which, as shown in Appendix \ref{app:tau_transform}, is a unbiased estimate of $|\bm{r}_k - \bm{w}_0|^2$ when $\bm{r}_k$ has unbiased independent entries, as expected by state evolution. $|\bm{\Psi F}^H|^2$ is the power spectrum of $\bm{\Psi}$, so has $1+3s$ unique rows; line 5 therefore requires $1+3s$ inner products. For fixed $s$ the complexity of VDAMP is therefore governed by $\bm{\Psi}$ and $\bm{F}$, whose fast implementations have complexity $O(N)$ and $O(N\log N)$ respectively. 

\subsection{Complex soft threshold tuning with SURE, line 6}

Selecting appropriate regularisation parameters such as $\lambda$ is a notable challenge in real-world compressed sensing applications. We present an approach to parameter-free compressed sensing reconstruction that leverages VDAMP's state evolution by applying Stein's Unbiased Risk Estimate (SURE) \cite{Stein1981}, building on work on AMP in \cite{Mousavi2013, Guo2015 , Bayati2013}. 

A strength of automatic parameter tuning via SURE is that it is possible to have a richer regularizer than would usually be feasible for a hand-tuned $f(\bm{x})$. In this work, the denoiser $\bm{g}(\bm{r}_{k}; \bm{\tau}_{k})$ was the complex soft thresholding operator with a subband-dependent threshold that is tuned automatically with SURE. In other words, SURE with an effective noise model given by $\bm{r}_{k} = \bm{w}_0 + \mathcal{CN}(\bm{0},  \mathrm{Diag}(\bm{\tau}_k) )$ was used to approximately solve
\begin{equation}
        \bm{g}(\bm{r}_{k}; \bm{\tau}_{k}) \approx \underset{ \bm{w}  \in \mathds{C}^N}{\operatorname{argmin}} \underset{\bm{\lambda}  \in \mathds{R}^N}{\operatorname{min}} \frac{1}{2} \|(\bm{w} - \bm{r}_{k})\oslash \sqrt{\bm{\tau}_{k}}\|^2_2  + \|\bm{\lambda} \odot \bm{w}\|_1, \label{eqn:gSURE}
\end{equation}
where $\oslash$ denotes entry-wise division, $\sqrt{\bm{\tau}_{k}}$ is the entry-wise square root of $\bm{\tau}_{k}$ and $\bm{\lambda}$ has the piecewise-constant structure of \eqref{eqn:vecalpha}. The possibility of using a scale-dependent thresholds is well known, such as in \cite{Daubechies2004}. We emphasize that \eqref{eqn:gSURE} is \textit{subband}-dependent rather than scale-dependent, enabling higher order, anisotropic signal modeling \cite{Vonesch2008, Bayram2010a}.  

Equation \eqref{eqn:gSURE} was solved using a procedure related to SureShrink \cite{Donoho1995} but for Gaussian noise that is complex and colored \cite{Johnstone1997}. Consider a vector $\bm{v}_0 \in \mathds{C}^{N_v}$ corrupted by white complex Gaussian noise: $\bm{v} = \bm{v}_0 + \mathcal{CN}(\bm{0}, \tau_v \mathds{1}_{N_v})$.  Let $\bm{d}(\bm{v}) = \bm{v} + \bm{h}(\bm{v})$ be an estimator of $\bm{v}_0$. SURE \cite{Stein1981} adapted to complex variables is 
\begin{equation}
cSURE(\bm{d}(\bm{v})) =  \|\bm{h}(\bm{v})\|^2_2  + N_v \tau_v [2\braket{\bm{\partial} (\bm{d}(\bm{v}))} - 1].
\label{eqn:cSURE_general}
\end{equation} 
cSURE is of interest to denoising problems because, as shown in Appendix \ref{app:cSURE}, it is an unbiased estimate of the risk $\mathds{E}\{\|\bm{d}(\bm{v}) - \bm{v}_0\|^2_2$\}. The optimal parameters of the denoiser $\bm{d}(\bm{v})$  can therefore be estimated by minimizing cSURE as a proxy for the true risk. Consider the case where $\bm{d}(\bm{v})$ is the complex soft thresholding operator $\bm{\eta}(\bm{v}; t)$ with threshold $t$, which acts component-wise as 
\begin{equation}
    \eta_j(v_j; t) := v_j \left(1 - \min\left\{\frac{t}{|v_j|},1\right\}\right). \label{eqn:soft_thr}
\end{equation}
The $j$th entry of  $\bm{\partial}(\bm{\eta}(\bm{v}; t))$ is
\begin{equation}
\partial_j ( \eta_j(v_j; t) ) = \begin{cases} \mbox{$0$,} & \mbox{if } |v_j|\leq t \\ 
\mbox{$1 - \frac{t}{2|v_j|}$, }  & \mbox{otherwise.} \end{cases} 
\end{equation}
By \eqref{eqn:cSURE_general}, an unbiased estimate of the risk of soft thresholding is therefore \cite{Donoho1995}
\begin{multline}
    cSURE(\bm{\eta}(\bm{v} ; t)) = (t^2 + 2\tau_v) \cdot \#\{ j : |v_j| > t\} -N_v \tau_v  \\
    +  \sum_{j: |v_j| \leq t}^{N_v}|v_j|^2 -    \sum_{j: |v_j| > t}^{N_v}t\tau_v/|v_j|.
    \label{eqn:cSURE}
\end{multline}
The optimal threshold for each subband can be estimated with
\begin{equation}
\hat{t} = \underset{t}{\operatorname{argmin}}(cSURE(\bm{\eta} (\bm{v} ; t)))  \label{eqn:that}
\end{equation}
by evaluating \eqref{eqn:cSURE} for trial thresholds $t = |v_1|, |v_2|,\ldots, |v_{N_v}|$. For large dimension $N_v$ one would expect by the law of large numbers that cSURE is close to the true risk, and for the threshold to be almost optimal. Since a larger number of decomposition scales $s$ give subbands with lower dimension, there is a trade-off between the size of $s$ and the quality of threshold selection with cSURE.    

SURE has previously been employed for parameter-free compressed sensing MRI in  \cite{Khare2012}, where FISTA was used with \eqref{eqn:gSURE} in place of the usual shrinkage step. This algorithm is herein referred to as SURE-IT, and is discussed in detail in Section \ref{sec:FISTA_based}. 

\subsection{Complex, colored Onsager correction, lines 7-9 \label{sec:VDAMP_ons_corr} }
 
Lines 7-9 of Algorithm \ref{alg:VDAMP} can be understood intuitively as follows: since lines 7-9 of Algorithm \ref{alg:OAMP} apply to white noise, and the effective noise of VDAMP is white within each subband, the Onsager correction must be applied subband-by-subband. For soft thresholding, the use of $\braket{\bm{\partial}(\bm{g}(\bm{r}_{k}; \bm{\tau}_{k}))}_\text{sband}$ in place of OAMP's $\braket{\bm{g}'(\bm{r}_{k}; \tau_{k})}$ leads the function formed by merging lines 7-9,
 \begin{equation}
    \tilde{\bm{g}}(\bm{r}_{k}; \bm{\tau}_{k}) = \bm{c}_k \odot (\bm{g}(\bm{r}_{k}; \bm{\tau}_{k}) - \bm{\alpha}_k \odot \bm{r}_{k}), \label{eqn:g_tilde}
 \end{equation}
to obey, for all subbands $b$,
\begin{align}
  \frac{1}{N_b} \sum_{j \in J_b}  \frac{\partial  \Re[\tilde{g}_j(\bm{r}_k; \bm{\tau}_{k})]}{\partial \Re[r_j]} \approx  \frac{1}{N_b} \sum_{j \in J_b} \frac{\partial  \Im[\tilde{g}_j(\bm{r}_k; \bm{\tau}_{k})]}{\partial \Im[r_j]} \approx 0, \label{eqn:VDAMP_div_free}
\end{align} 
which is a colored aliasing version of OAMP's divergence-free condition, \eqref{eqn:div_free}, applied to both real and imaginary parts. The Onsager correction employed here is not the only choice that leads to a $\tilde{\bm{g}}(\bm{r}_{k}; \bm{\tau}_{k})$ that satisfies \eqref{eqn:VDAMP_div_free}\cite{maleki2013asymptotic}, however, we have found that this particular choice performs well.   

Like the scalar $c_k$ in Algorithm \ref{alg:OAMP}, the $\bm{c}_k$ updated in line 8 of Algorithm \ref{alg:VDAMP} is not constrained by \eqref{eqn:VDAMP_div_free}, except to have  the piecewise-constant structure of \eqref{eqn:vecalpha}. In the experiments in this work, two possibilities for the $\bm{c}_k$ update are considered. First,
\begin{equation}
    \bm{c}^{\alpha}_k = \bm{1}_N \oslash (\bm{1}_N-\bm{\alpha}_k). \label{eqn:Cvamp}
\end{equation}
In this case, lines 6-9 of Algorithm \ref{alg:VDAMP} are a colored version of the `denoising' phase of VAMP when written in LMMSE form (see \cite{Rangan2019}, Algorithm 3). VDAMP with $\bm{c}_k$ updated with \eqref{eqn:Cvamp} is herein referred to as VDAMP-$\alpha$.  

Secondly, as in \cite{Xue2016}, we suggest using cSURE for a second time to estimate the $\bm{c}_k$ that minimizes the mean squared error of $\widetilde{\bm{r}}_k$:  
\begin{equation}
    \bm{c}_k^{SURE} \approx \argmin_{\bm{c}} \| \tilde{\bm{g}}(\bm{r}_k, \bm{\tau}_k) - \bm{w}_0 \|^2_2.
    \label{eqn:Copt}
\end{equation}
Since by \eqref{eqn:VDAMP_div_free} $\braket{\bm{\partial} (\widetilde{\bm{g}}(\bm{r}_{k}; \bm{\tau}_{k}))}_\mathrm{sband} \approx 0$, optimizing  \eqref{eqn:cSURE_general} reduces to a series of $\ell_2$ minimization problems with a closed-form solution, so that for the $b$th subband
\begin{subequations}
\label{eqn:Csure}
\begin{align}
     c_{k,b}^{SURE} 
     &= \argmin_{c} \| c (\bm{g}_b(\bm{r}_{k} ; \bm{\tau}_{k}) -  \alpha_{k,b} \bm{r}_{k,b} ) - \bm{r}_{k,b}\|^2_2 \\
     &= \frac{\bm{r}_{k,b}^H (\bm{g}_b(\bm{r}_{k} ; \bm{\tau}_{k}) -  \alpha_{k,b} \bm{r}_{k,b} )}{\|\bm{g}_b(\bm{r}_{k} ; \bm{\tau}_{k}) -  \alpha_{k, b} \bm{r}_{k,b}\|^2_2}. 
\end{align}
\end{subequations}
Vector $\bm{c}_k^{SURE}$ is formed from the $c_{k,b}^{SURE}$ so that it has the structure of \eqref{eqn:vecalpha}.  VDAMP with $\bm{c}_k$ updated with $\bm{c}_k^{SURE}$ is herein referred to as VDAMP-S.  

As for OAMP \cite{Ma2017}, we do not claim that the either of the two $\bm{c}_k$ updates suggested here are necessarily optimal. Instead, we show in Section \ref{sec:NumExp} that both updates lead to aliasing consistent with state evolution, and empirically evaluate their performance, which we observe to converge to a similar or lower NMSE as other 
algorithms, but with approximately a tenth the time to convergence.

 
 \section{Numerical experiments\label{sec:NumExp}} This section illustrates the performance of VDAMP-$\alpha$ and VDAMP-S compared with the Fast Iterative Shrinking-Thresholding Algorithm (FISTA) algorithm \cite{Daubechies2004,Beck2009}, and two FISTA-based methods wth subband-dependent thresholding \cite{Bayram2010a, Khare2012}. We also present empirical evidence for VDAMP's state evolution. All experiments were conducted in MATLAB 9.4 and can be reproduced with code  online, available at \url{https://github.com/charlesmillard/VDAMP}.

\subsection{Description of comparative FISTA-based algorithms \label{sec:FISTA_based}}

\begin{algorithm}[t]
\caption{FISTA \cite{Beck2009}, S-FISTA \cite{Bayram2010a} and SURE-IT \cite{ Khare2012}  for Fourier sampled images  \label{alg:FISTA}}
\textbf{Require:} Sampling set $\Omega$, wavelet transform $\bm{\Psi}$, per-subband weighting $\bm{W}$ set to $\bm{W} = \mathds{1}_N$ for FISTA and SURE-IT or calculated with \eqref{eqn:s-fista_weight} for S-FISTA, sparse weighting $\lambda$ for FISTA and S-FISTA, measurements $\bm{y}$, number of iterations $K_{it}$.
\begin{algorithmic}[1]
\STATE Set $\widetilde{\bm{r}}_{0} = \bm{0}$, $\hat{\bm{w}}_{-1} = \bm{0}$ and  $h_{-1} = 1$ 
\FOR {$k =0,1,\ldots, K_{it}-1$} 
\STATE $\bm{z}_k = \bm{y} -  \bm{F} \bm{\Psi}^H\widetilde{\bm{r}}_{k}$ 
\STATE $\bm{r}_{k} = \widetilde{\bm{r}}_{k} + \bm{W}^{-1} \bm{\Psi F}^H \bm{M}_\Omega \bm{z}_k$ 
\STATE Update $\tau_k$ 
\IF{SURE-IT}
	\STATE $\hat{\bm{w}}_k = \bm{g} (\bm{r}_{k}; {\tau_k \bm{1} })$ 
\ELSE
	\STATE $\hat{\bm{w}}_k = \bm{\eta} (\bm{r}_{k}; {\tau_k \lambda \bm{W}^{-1} \bm{1} })$ 
\ENDIF
\STATE $h_k = (1 + \sqrt{1+4h_{k-1}^2})/2$
\STATE $\widetilde{\bm{r}}_{k+1} = \hat{\bm{w}}_k + (h_{k-1} - 1)(\hat{\bm{w}}_k - \hat{\bm{w}}_{k-1})/h_k$
\ENDFOR
\RETURN $\hat{\bm{x}} = \bm{\Psi}^H\hat{\bm{w}}_k + \bm{F}^H (\bm{y} - \bm{M}_\Omega\bm{F} \bm{\Psi}^H \hat{\bm{w}}_k)$ 
\end{algorithmic}
\end{algorithm}

The three versions of FISTA \cite{Beck2009, Bayram2010a, Khare2012} considered in this work are summarized in Algorithm \ref{alg:FISTA}.  In line 7, $\bm{g} (\bm{r}_{k}; {\tau_k \bm{1} })$ refers to the subband-dependent thresholding function with thresholds tuned by cSURE, stated in \eqref{eqn:gSURE}, with a scalar aliasing model $\tau_k$. In line 9, $\bm{\eta} (\bm{r}_{k}; {\tau_k\lambda  \bm{\Lambda}^{-1} \bm{1} })$ is the soft thresholding function from \eqref{eqn:soft_thr} with entry-wise threshold $\tau_k \lambda  \bm{\Lambda}^{-1} \bm{1}$. As in \eqref{eqn:prox_denoi}, the threshold employed here is weighted by the variance estimate $\tau_k$, causing the threshold to decrease over iterations, which, as in \cite{Vasanawala2011}, was found to significantly reduce the time to convergence. Line 14 is an unweighted gradient descent step, as in line 11 of Algorithm \ref{alg:VDAMP}, which outputs an estimate $\hat{\bm{x}}$ with high data fidelity. This was suggested in \cite{Vasanawala2011}, and we found that this output had a lower reconstruction error than $\bm{\Psi}^H \hat{\bm{w}}_k$. The variations described in Algorithm \ref{alg:FISTA}, discussed in detail below, are referred to in this work as \textit{FISTA}, \textit{S-FISTA} \cite{Bayram2010a} and \textit{SURE-IT} \cite{ Khare2012}.

\textit{FISTA} refers to Algorithm \ref{alg:FISTA} with $\bm{W} = \mathds{1}_N$, so that a global threshold $\lambda \tau_k$ is applied in line 9. Despite not discriminating between subbands, this version of FISTA was considered because it is widely-used in MRI, and we found that it performed well compared with the subband-dependent algorithms, so is an instructive baseline.  FISTA with a hand-tuned subband-dependent $\lambda$ was not considered as it is not feasible in practice.

\textit{S-FISTA} refers to Algorithm \ref{alg:FISTA} with a diagonal weight matrix $\bm{W}$ that has one unique entry per wavelet subband, so that $\mathrm{diag}(\bm{W})$ has the piecewise-constant structure of \eqref{eqn:vecalpha}. In \cite{Bayram2010a}, a method was suggested for selecting the weight for the $b$th subband, which we denote as $w_b$. Let $\bm{\Phi}_b$ be the block of $\bm{M}_\Omega \bm{F} \bm{\Psi}^H$ corresponding to the $b$th subband. In \cite{Bayram2010a}, it is shown that a choice of $w_b$ that satisfies
\begin{equation}
\frac{1}{w_b} > \sum_{b' = 1}^{3s+1}  \sqrt{\lambda_{\text{max}}( \bm{\Phi}_{b'}^H\bm{\Phi}^{ }_{b^{ }} \bm{\Phi}_b^H \bm{\Phi}^{ }_{b'})}, \label{eqn:s-fista_weight}
\end{equation}
where $\lambda_{\text{max}}(\cdot)$ is the largest eigenvalue, ensures that $\bm{W} - \bm{\Phi}^H\bm{\Phi}$ is a positive operator, and therefore guarantees that the algorithm converges: see \cite{Bayram2010a} for details. Following \cite{Guerquin-Kern2011}, we computed the $w_b$ by calculating $\lambda_{\text{max}}$ once per $\Omega$ for all $b$ and $b'$ using the power iteration method. In contrast with VDAMP, S-FISTA's per-subband weighting is fixed across all iterations, and depends only on the sensing matrix and wavelet family, and not on the per-iteration signal estimate.

\textit{SURE-IT} refers to Algorithm \ref{alg:FISTA} with $\bm{W} = \mathds{1}_N$ and a subband-dependent, automatically tuned denoiser $\bm{g} (\bm{r}_{k}; {\tau_k \bm{1}})$, as in \cite{ Khare2012}. Although SURE-IT's thresholding is iteration-dependent, the aliasing of its $\bm{r}_k$ is highly non-Gaussian, so deviates from a proper theoretical basis for using SURE and does not give near-optimal threshold selection. Further, unlike VDAMP, the aliasing of SURE-IT is modeled by a scalar $\tau_k$. 

In the following experiments, the scalar sparse weighting $\lambda$ of FISTA and S-FISTA was tuned with an exhaustive search so that the mean-squared error  was  minimized at $k=100$. Since the threshold was weighted by $\tau_k$, we tuned $\lambda$ separately for FISTA and S-FISTA. The variance estimate $\tau_k$  on line 5 of Algorithm \ref{alg:FISTA} was updated using the ground truth: $\tau_k =  \|\bm{r}_k - \bm{w}_0\|^2_2/N$. 
 
\begin{figure}[t!]
\hfill
 \subfloat[$256\times 256$ Brain]{%
    \includegraphics[height=\threeH \columnwidth]{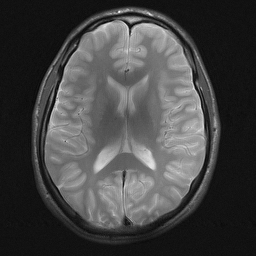}}
    \hfill
\subfloat[$208\times 416$ Cardiac]{%
      \includegraphics[height = \threeH \columnwidth]{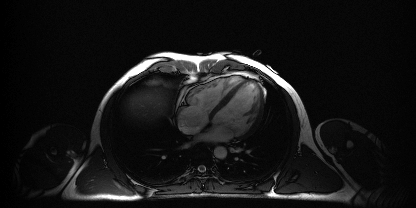}}
     \caption{\label{fig:test_images} MRI images used to evaluate the performance of VDAMP.}
     \hfill
\end{figure}   
 
\subsection{Experimental method}

We considered the reconstruction of 8 test images: the Shepp-Logan shown in Fig. \ref{fig:shepp_logan}, a brain and a cardiac MRI image, shown in Fig. \ref{fig:test_images}, and 5 standard test images:  Barbara, Boat, Cameraman, House and Peppers, downloaded from \cite{Makinen2020}. In all cases the undersampled data was artificially corrupted with complex Gaussian noise so that $\|\bm{ x}_0\|^2_2/N\sigma_\varepsilon^2 = 40\mathrm{dB}$. We assumed that $\sigma_\varepsilon^2$ was known a priori. For simplicity $\bm{\Psi}$ was chosen as the Haar wavelet with s=4 scales.

\begin{figure*}[!h]
\centering
   \subfloat[\label{fig:FISTAx} FISTA, NMSE -14.2dB]{%
      \includegraphics[width=0.65\columnwidth]{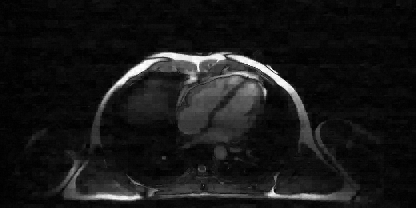}}
\hspace{\fill} 
   \subfloat[\label{fig:SUREITx} S-FISTA, NMSE -11.3dB]{%
      \includegraphics[width=0.65\columnwidth]{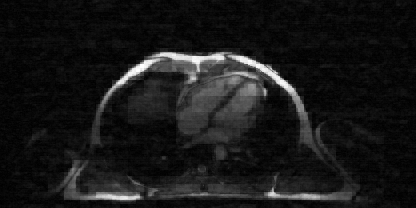}}
      \hspace{\fill} 
         \subfloat[\label{fig:SUREITx} SURE-IT, NMSE -12.2dB]{%
      \includegraphics[width=0.65\columnwidth]{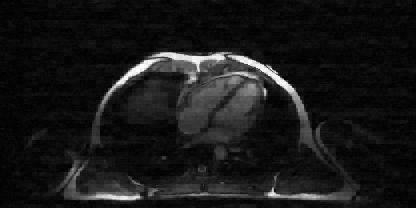}}
      \hspace{\fill}
      \\ \centering
	\subfloat[\label{fig:VDAMPalpha} VDAMP-$\alpha$, NMSE -17.9dB]{%
      \includegraphics[width=0.65\columnwidth]{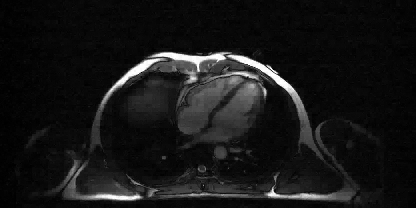}}
	\hspace{\fill} 
   \subfloat[\label{fig:VDAMPS} VDAMP-S, NMSE -18.0dB]{%
      \includegraphics[width=0.65\columnwidth]{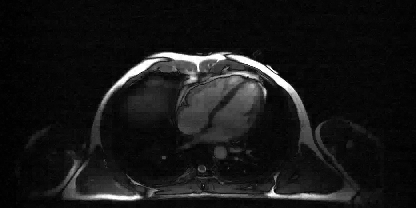}} 
	\hspace{\fill} 
   \subfloat[\label{fig:mask} $N/n = 4$ sampling set $\Omega$]{%
      \includegraphics[width=0.65\columnwidth]{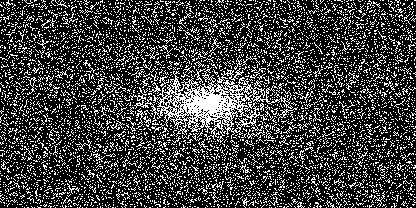}} 
      \caption{\label{fig:cardiac_example} Reconstructions of the Cardiac image undersampled at $N/n = 4$, shown at $k=10$.}
\end{figure*}

A variety of variable density sampling schemes for MRI have been suggested \cite{Tsai2000, wangVDS, Lustig2007, Chauffert2013,  Lazarus2019, Vasanawala2011}, including some with recovery guarantees  \cite{Krahmer2014, Adcock2017}. In the experiments presented in this section, we generated $\Omega$ using the variable density sampling function from the Sparse MRI package\footnote{available at {https://people.eecs.berkeley.edu/\texttildelow mlustig/Software.html}}. The focus of this work is on the reconstruction algorithm, not on the sampling scheme, and we do not claim that this scheme is necessarily the best choice, only that it an instructive example because it common in MRI and known to perform well in practice \cite{Lustig2007}. The $\Omega$ generated with this package for the $N/n = 4$ cardiac image is shown in Fig.  \ref{fig:mask}.  We chose variable density schemes $\bm{P}$ with $\mathds{E}_\Omega\{ |\Omega| \} = n$ so that the acceleration factors $N/n = 4, 6, 8$ were used, except for the Shepp-Logan, where we used an increased acceleration of $N/n = 8, 10, 12$.   All algorithms were initialized with a vector of zeros and run for $K_{it} = 500$ iterations, except for the Shepp-Logan, which was found to require more iterations, so was run until $K_{it} = 1000$.  

\begin{table*}[]
\centering
\begin{tabular}{crrrrrrrrrrrr}
\hline
\multicolumn{1}{|c||}{Image}          & \multicolumn{3}{c||}{\textsc{Shepp-Logan}}                                                                             & \multicolumn{3}{c||}{\textsc{Brain}}                                                                                    & \multicolumn{3}{c||}{\textsc{Cardiac}}                                                                                 & \multicolumn{3}{c|}{\textsc{Barbara}}                                                                               \\ \hline
\multicolumn{1}{|c||}{$n/N$}          & \multicolumn{1}{c|}{8}             & \multicolumn{1}{c|}{10}            & \multicolumn{1}{c||}{12}            & \multicolumn{1}{c|}{4}             & \multicolumn{1}{c|}{6}             & \multicolumn{1}{c||}{8}             & \multicolumn{1}{c|}{4}             & \multicolumn{1}{c|}{6}             & \multicolumn{1}{c||}{8}             & \multicolumn{1}{c|}{4}             & \multicolumn{1}{c|}{6}             & \multicolumn{1}{c|}{8}             \\  \hhline{|=#=|=|=#=|=|=#=|=|=#=|=|=|}
\multicolumn{1}{|c||}{FISTA}          & \multicolumn{1}{r|}{57.99}         & \multicolumn{1}{r|}{69.55}         & \multicolumn{1}{r||}{76.64}         & \multicolumn{1}{r|}{1.40}          & \multicolumn{1}{r|}{1.53}          & \multicolumn{1}{r||}{1.68}          & \multicolumn{1}{r|}{3.37}          & \multicolumn{1}{r|}{3.45}          & \multicolumn{1}{r||}{3.20}          & \multicolumn{1}{r|}{1.01}          & \multicolumn{1}{r|}{1.65}          & \multicolumn{1}{r|}{3.18}          \\ \hline
\multicolumn{1}{|c||}{S-FISTA}        & \multicolumn{1}{r|}{79.91}         & \multicolumn{1}{r|}{80.62}         & \multicolumn{1}{r||}{81.13}         & \multicolumn{1}{r|}{4.22}          & \multicolumn{1}{r|}{4.93}          & \multicolumn{1}{r||}{4.79}          & \multicolumn{1}{r|}{10.41}         & \multicolumn{1}{r|}{8.53}          & \multicolumn{1}{r||}{9.66}          & \multicolumn{1}{r|}{6.00}          & \multicolumn{1}{r|}{6.78}          & \multicolumn{1}{r|}{10.10}         \\ \hline
\multicolumn{1}{|c||}{SURE-IT}        & \multicolumn{1}{r|}{8.60}          & \multicolumn{1}{r|}{4.45}          & \multicolumn{1}{r||}{5.25}          & \multicolumn{1}{r|}{0.91}          & \multicolumn{1}{r|}{0.90}          & \multicolumn{1}{r||}{1.08}          & \multicolumn{1}{r|}{1.84}          & \multicolumn{1}{r|}{20.77}         & \multicolumn{1}{r||}{1.67}          & \multicolumn{1}{r|}{1.05}          & \multicolumn{1}{r|}{1.15}          & \multicolumn{1}{r|}{4.50}          \\ \hline
\multicolumn{1}{|c||}{VDAMP-$\alpha$} & \multicolumn{1}{r|}{2.99}          & \multicolumn{1}{r|}{4.41}          & \multicolumn{1}{r||}{\textbf{0.96}} & \multicolumn{1}{r|}{\textbf{0.16}} & \multicolumn{1}{r|}{\textbf{0.13}} & \multicolumn{1}{r||}{0.14}          & \multicolumn{1}{r|}{\textbf{0.39}} & \multicolumn{1}{r|}{0.38}          & \multicolumn{1}{r||}{\textbf{0.24}} & \multicolumn{1}{r|}{\textbf{0.18}} & \multicolumn{1}{r|}{\textbf{0.19}} & \multicolumn{1}{r|}{0.43}          \\ \hline
\multicolumn{1}{|c||}{VDAMP-S}        & \multicolumn{1}{r|}{\textbf{2.21}} & \multicolumn{1}{r|}{\textbf{2.37}} & \multicolumn{1}{r||}{4.70}          & \multicolumn{1}{r|}{0.28}          & \multicolumn{1}{r|}{\textbf{0.13}} & \multicolumn{1}{r||}{\textbf{0.11}} & \multicolumn{1}{r|}{0.43}          & \multicolumn{1}{r|}{\textbf{0.26}} & \multicolumn{1}{r||}{1.28}          & \multicolumn{1}{r|}{0.56}          & \multicolumn{1}{r|}{0.33}          & \multicolumn{1}{r|}{\textbf{0.31}} \\ \hline
\multicolumn{1}{l}{}                 & \multicolumn{1}{l}{}               & \multicolumn{1}{l}{}               & \multicolumn{1}{l}{}               & \multicolumn{1}{l}{}               & \multicolumn{1}{l}{}               & \multicolumn{1}{l}{}               & \multicolumn{1}{l}{}               & \multicolumn{1}{l}{}               & \multicolumn{1}{l}{}               & \multicolumn{1}{l}{}               & \multicolumn{1}{l}{}               & \multicolumn{1}{l}{}               \\ \hline
\multicolumn{1}{|c||}{Image}          & \multicolumn{3}{c||}{\textsc{Boat}}                                                                                 & \multicolumn{3}{c||}{\textsc{Cameraman}}                                                                                   & \multicolumn{3}{c||}{\textsc{House}}                                                                                   & \multicolumn{3}{c|}{\textsc{Peppers}}                                                                                 \\ \hline
\multicolumn{1}{|c||}{$n/N$}          & \multicolumn{1}{c|}{4}             & \multicolumn{1}{c|}{6}             & \multicolumn{1}{c||}{8}             & \multicolumn{1}{c|}{4}             & \multicolumn{1}{c|}{6}             & \multicolumn{1}{c||}{8}             & \multicolumn{1}{c|}{4}             & \multicolumn{1}{c|}{6}             & \multicolumn{1}{c||}{8}             & \multicolumn{1}{c|}{4}             & \multicolumn{1}{c|}{6}             & \multicolumn{1}{c|}{8}             \\ \hhline{|=#=|=|=#=|=|=#=|=|=#=|=|=|}
\multicolumn{1}{|c||}{FISTA}          & \multicolumn{1}{r|}{2.80}          & \multicolumn{1}{r|}{4.06}          & \multicolumn{1}{r||}{3.08}          & \multicolumn{1}{r|}{1.56}          & \multicolumn{1}{r|}{1.58}          & \multicolumn{1}{r||}{1.22}          & \multicolumn{1}{r|}{1.52}          & \multicolumn{1}{r|}{1.97}          & \multicolumn{1}{r||}{1.94}          & \multicolumn{1}{r|}{1.43}          & \multicolumn{1}{r|}{1.66}          & \multicolumn{1}{r|}{1.94}          \\ \hline
\multicolumn{1}{|c||}{S-FISTA}        & \multicolumn{1}{r|}{10.44}         & \multicolumn{1}{r|}{11.71}         & \multicolumn{1}{r||}{8.13}          & \multicolumn{1}{r|}{5.15}          & \multicolumn{1}{r|}{4.94}          & \multicolumn{1}{r||}{4.20}          & \multicolumn{1}{r|}{4.44}          & \multicolumn{1}{r|}{5.08}          & \multicolumn{1}{r||}{4.44}          & \multicolumn{1}{r|}{5.24}          & \multicolumn{1}{r|}{5.29}          & \multicolumn{1}{r|}{5.57}          \\ \hline
\multicolumn{1}{|c||}{SURE-IT}        & \multicolumn{1}{r|}{2.31}          & \multicolumn{1}{r|}{2.73}          & \multicolumn{1}{r||}{2.36}          & \multicolumn{1}{r|}{1.15}          & \multicolumn{1}{r|}{0.94}          & \multicolumn{1}{r||}{0.81}          & \multicolumn{1}{r|}{2.27}          & \multicolumn{1}{r|}{2.09}          & \multicolumn{1}{r||}{3.03}          & \multicolumn{1}{r|}{0.73}          & \multicolumn{1}{r|}{0.19}          & \multicolumn{1}{r|}{0.15}          \\ \hline
\multicolumn{1}{|c||}{VDAMP-$\alpha$} & \multicolumn{1}{r|}{\textbf{0.44}} & \multicolumn{1}{r|}{\textbf{0.42}} & \multicolumn{1}{r||}{\textbf{0.31}} & \multicolumn{1}{r|}{\textbf{0.19}} & \multicolumn{1}{r|}{\textbf{0.13}} & \multicolumn{1}{r||}{0.73}          & \multicolumn{1}{r|}{0.17}          & \multicolumn{1}{r|}{0.17}          & \multicolumn{1}{r||}{0.14}          & \multicolumn{1}{r|}{0.14}          & \multicolumn{1}{r|}{\textbf{0.09}} & \multicolumn{1}{r|}{\textbf{0.09}} \\ \hline
\multicolumn{1}{|c||}{VDAMP-S}        & \multicolumn{1}{r|}{0.70}          & \multicolumn{1}{r|}{3.62}          & \multicolumn{1}{r||}{2.23}          & \multicolumn{1}{r|}{0.99}          & \multicolumn{1}{r|}{0.68}          & \multicolumn{1}{r||}{\textbf{0.06}} & \multicolumn{1}{r|}{\textbf{0.14}} & \multicolumn{1}{r|}{\textbf{0.09}} & \multicolumn{1}{r||}{\textbf{0.09}} & \multicolumn{1}{r|}{\textbf{0.10}} & \multicolumn{1}{r|}{0.10}          & \multicolumn{1}{r|}{\textbf{0.09}} \\ \hline
\end{tabular}
\caption{\label{tab:conv_time}  Convergence time in seconds. The shortest time is highlighted in bold.}
\vspace{1cm}
\centering
\begin{tabular}{crrrrrrrrrrrr}
\hline
\multicolumn{1}{|c||}{Image}          & \multicolumn{3}{c||}{\textsc{Shepp-Logan}}                                                                                & \multicolumn{3}{c||}{\textsc{Brain}}                                                                                       & \multicolumn{3}{c||}{\textsc{Cardiac}}                                                                                    & \multicolumn{3}{c|}{\textsc{Barbara}}                                                                                  \\ \hline
\multicolumn{1}{|c||}{$n/N$}          & \multicolumn{1}{c|}{8}              & \multicolumn{1}{c|}{10}             & \multicolumn{1}{c||}{12}             & \multicolumn{1}{c|}{4}              & \multicolumn{1}{c|}{6}              & \multicolumn{1}{c||}{8}              & \multicolumn{1}{c|}{4}              & \multicolumn{1}{c|}{6}              & \multicolumn{1}{c||}{8}              & \multicolumn{1}{c|}{4}              & \multicolumn{1}{c|}{6}              & \multicolumn{1}{c|}{8}              \\ \hhline{|=#=|=|=#=|=|=#=|=|=#=|=|=|}
\multicolumn{1}{|c||}{FISTA}          & \multicolumn{1}{r|}{-36.2}          & \multicolumn{1}{r|}{-31.9}          & \multicolumn{1}{r||}{-25.7}          & \multicolumn{1}{r|}{-20.4}          & \multicolumn{1}{r|}{-18.2}          & \multicolumn{1}{r||}{-17.1}          & \multicolumn{1}{r|}{-17.9}          & \multicolumn{1}{r|}{\textbf{-14.5}} & \multicolumn{1}{r||}{\textbf{-13.0}} & \multicolumn{1}{r|}{\textbf{-17.5}} & \multicolumn{1}{r|}{\textbf{-16.3}} & \multicolumn{1}{r|}{-15.5}          \\ \hline
\multicolumn{1}{|c||}{S-FISTA}        & \multicolumn{1}{r|}{-33.2}          & \multicolumn{1}{r|}{-28.1}          & \multicolumn{1}{r||}{-23.5}          & \multicolumn{1}{r|}{-20.3}          & \multicolumn{1}{r|}{-18.1}          & \multicolumn{1}{r||}{-17.1}          & \multicolumn{1}{r|}{-17.7}          & \multicolumn{1}{r|}{-14.3}          & \multicolumn{1}{r||}{-12.7}          & \multicolumn{1}{r|}{\textbf{-17.5}} & \multicolumn{1}{r|}{\textbf{-16.3}} & \multicolumn{1}{r|}{-15.5}          \\ \hline
\multicolumn{1}{|c||}{SURE-IT}        & \multicolumn{1}{r|}{-18.7}          & \multicolumn{1}{r|}{-13.3}          & \multicolumn{1}{r||}{-13.1}          & \multicolumn{1}{r|}{-18.9}          & \multicolumn{1}{r|}{-17.3}          & \multicolumn{1}{r||}{-16.6}          & \multicolumn{1}{r|}{-14.9}          & \multicolumn{1}{r|}{-12.9}           & \multicolumn{1}{r||}{-10.5}          & \multicolumn{1}{r|}{-17.0}          & \multicolumn{1}{r|}{-16.1}          & \multicolumn{1}{r|}{-15.5}          \\ \hline
\multicolumn{1}{|c||}{VDAMP-$\alpha$} & \multicolumn{1}{r|}{-35.1}          & \multicolumn{1}{r|}{-29.4}          & \multicolumn{1}{r||}{-20.5}          & \multicolumn{1}{r|}{\textbf{-20.7}}          & \multicolumn{1}{r|}{\textbf{-18.5}}          & \multicolumn{1}{r||}{\textbf{-17.5}} & \multicolumn{1}{r|}{-17.9}          & \multicolumn{1}{r|}{\textbf{-14.5}} & \multicolumn{1}{r||}{-12.9}          & \multicolumn{1}{r|}{\textbf{-17.5}} & \multicolumn{1}{r|}{\textbf{-16.3}} & \multicolumn{1}{r|}{\textbf{-15.6}} \\ \hline
\multicolumn{1}{|c||}{VDAMP-S}        & \multicolumn{1}{r|}{\textbf{-38.1}} & \multicolumn{1}{r|}{\textbf{-34.9}} & \multicolumn{1}{r||}{\textbf{-30.8}} & \multicolumn{1}{r|}{-20.6} & \multicolumn{1}{r|}{-18.4} & \multicolumn{1}{r||}{-17.4}          & \multicolumn{1}{r|}{\textbf{-18.0}} & \multicolumn{1}{r|}{-14.3}          & \multicolumn{1}{r||}{-12.4}          & \multicolumn{1}{r|}{-16.8}          & \multicolumn{1}{r|}{-15.4}          & \multicolumn{1}{r|}{-14.1}          \\ \hline   
\multicolumn{1}{l}{}                 & \multicolumn{1}{l}{}                & \multicolumn{1}{l}{}                & \multicolumn{1}{l}{}                & \multicolumn{1}{l}{}                & \multicolumn{1}{l}{}                & \multicolumn{1}{l}{}                & \multicolumn{1}{l}{}                & \multicolumn{1}{l}{}                & \multicolumn{1}{l}{}                & \multicolumn{1}{l}{}                & \multicolumn{1}{l}{}                & \multicolumn{1}{l}{}                \\ \hline
\multicolumn{1}{|c||}{Image}          & \multicolumn{3}{c||}{\textsc{Boat}}                                                                                    & \multicolumn{3}{c||}{\textsc{Cameraman}}                                                                                      & \multicolumn{3}{c||}{\textsc{House}}                                                                                      & \multicolumn{3}{c|}{\textsc{Peppers}}                                                                                    \\ \hline
\multicolumn{1}{|c||}{$n/N$}          & \multicolumn{1}{c|}{4}              & \multicolumn{1}{c|}{6}              & \multicolumn{1}{c||}{8}              & \multicolumn{1}{c|}{4}              & \multicolumn{1}{c|}{6}              & \multicolumn{1}{c||}{8}              & \multicolumn{1}{c|}{4}              & \multicolumn{1}{c|}{6}              & \multicolumn{1}{c||}{8}              & \multicolumn{1}{c|}{4}              & \multicolumn{1}{c|}{6}              & \multicolumn{1}{c|}{8}              \\ \hhline{|=#=|=|=#=|=|=#=|=|=#=|=|=|}
\multicolumn{1}{|c||}{FISTA}          & \multicolumn{1}{r|}{-21.4}          & \multicolumn{1}{r|}{-19.6}          & \multicolumn{1}{r||}{-18.6}          & \multicolumn{1}{r|}{\textbf{-20.9}} & \multicolumn{1}{r|}{\textbf{-18.4}} & \multicolumn{1}{r||}{\textbf{-17.1}} & \multicolumn{1}{r|}{\textbf{-25.4}} & \multicolumn{1}{r|}{-22.8}          & \multicolumn{1}{r||}{\textbf{-21.6}} & \multicolumn{1}{r|}{\textbf{-20.0}} & \multicolumn{1}{r|}{-17.5}          & \multicolumn{1}{r|}{-16.2}          \\ \hline
\multicolumn{1}{|c||}{S-FISTA}        & \multicolumn{1}{r|}{-21.4}          & \multicolumn{1}{r|}{-19.5}          & \multicolumn{1}{r||}{-18.5}          & \multicolumn{1}{r|}{-20.7}          & \multicolumn{1}{r|}{-18.3}          & \multicolumn{1}{r||}{-17.0}          & \multicolumn{1}{r|}{-25.3}          & \multicolumn{1}{r|}{-22.7}          & \multicolumn{1}{r||}{-21.4}          & \multicolumn{1}{r|}{-19.9}          & \multicolumn{1}{r|}{-17.5}          & \multicolumn{1}{r|}{-16.1}          \\ \hline
\multicolumn{1}{|c||}{SURE-IT}        & \multicolumn{1}{r|}{-20.3}          & \multicolumn{1}{r|}{-18.9}          & \multicolumn{1}{r||}{-18.3}          & \multicolumn{1}{r|}{-18.7}          & \multicolumn{1}{r|}{-16.7}          & \multicolumn{1}{r||}{-15.9}          & \multicolumn{1}{r|}{-23.3}          & \multicolumn{1}{r|}{-21.4}          & \multicolumn{1}{r||}{-20.6}          & \multicolumn{1}{r|}{-18.5}          & \multicolumn{1}{r|}{-16.0}          & \multicolumn{1}{r|}{-15.3}          \\ \hline
\multicolumn{1}{|c||}{VDAMP-$\alpha$} & \multicolumn{1}{r|}{\textbf{-21.6}} & \multicolumn{1}{r|}{\textbf{-19.9}} & \multicolumn{1}{r||}{\textbf{-18.9}} & \multicolumn{1}{r|}{-20.8}          & \multicolumn{1}{r|}{-18.2}          & \multicolumn{1}{r||}{-16.0}          & \multicolumn{1}{r|}{\textbf{-25.4}} & \multicolumn{1}{r|}{\textbf{-22.9}} & \multicolumn{1}{r||}{-21.5}          & \multicolumn{1}{r|}{-19.9}          & \multicolumn{1}{r|}{\textbf{-17.7}} & \multicolumn{1}{r|}{\textbf{-16.7}} \\ \hline
\multicolumn{1}{|c||}{VDAMP-S}        & \multicolumn{1}{r|}{-19.5}          & \multicolumn{1}{r|}{-16.7}          & \multicolumn{1}{r||}{-15.6}          & \multicolumn{1}{r|}{-18.7}          & \multicolumn{1}{r|}{-16.1}          & \multicolumn{1}{r||}{-15.5}          & \multicolumn{1}{r|}{-21.5}          & \multicolumn{1}{r|}{-18.8}          & \multicolumn{1}{r||}{-17.8}          & \multicolumn{1}{r|}{-18.0}          & \multicolumn{1}{r|}{-15.7}          & \multicolumn{1}{r|}{-14.3}          \\ \hline
\end{tabular}
\caption{\label{tab:NMSE} Reconstruction NMSE in dB at $K_{it}$ for all 8 test images at different undersampling factors. The lowest NMSE is highlighted in bold.}
\vspace{1cm}
\centering
\begin{tabular}{crrrrrrrrrrrr}
\hline
\multicolumn{1}{|c||}{Image}          & \multicolumn{3}{c||}{\textsc{Shepp-Logan}}                                                                               & \multicolumn{3}{c||}{\textsc{Brain}}                                                                                      & \multicolumn{3}{c||}{\textsc{Cardiac}}                                                                                  & \multicolumn{3}{c|}{\textsc{Barbara}}                                                                                 \\ \hline
\multicolumn{1}{|c||}{$n/N$}          & \multicolumn{1}{c|}{8}              & \multicolumn{1}{c|}{10}            & \multicolumn{1}{c||}{12}             & \multicolumn{1}{c|}{4}             & \multicolumn{1}{c|}{6}              & \multicolumn{1}{c||}{8}              & \multicolumn{1}{c|}{4}              & \multicolumn{1}{c|}{6}             & \multicolumn{1}{c||}{8}             & \multicolumn{1}{c|}{4}              & \multicolumn{1}{c|}{6}             & \multicolumn{1}{c|}{8}              \\ \hhline{|=#=|=|=#=|=|=#=|=|=#=|=|=|}
\multicolumn{1}{|c||}{FISTA}          & \multicolumn{1}{r|}{27.99}         & \multicolumn{1}{r|}{32.57}         & \multicolumn{1}{r||}{38.01}         & \multicolumn{1}{r|}{2.86}           & \multicolumn{1}{r|}{4.43}           & \multicolumn{1}{r||}{5.76}           & \multicolumn{1}{r|}{3.94}           & \multicolumn{1}{r|}{6.35}           & \multicolumn{1}{r||}{9.55}           & \multicolumn{1}{r|}{2.10}           & \multicolumn{1}{r|}{2.75}          & \multicolumn{1}{r|}{3.09}          \\ \hline
\multicolumn{1}{|c||}{S-FISTA}        & \multicolumn{1}{r|}{43.69}         & \multicolumn{1}{r|}{45.63}         & \multicolumn{1}{r||}{46.83}         & \multicolumn{1}{r|}{9.58}           & \multicolumn{1}{r|}{11.35}          & \multicolumn{1}{r||}{11.97}          & \multicolumn{1}{r|}{20.46}          & \multicolumn{1}{r|}{23.56}          & \multicolumn{1}{r||}{25.42}          & \multicolumn{1}{r|}{3.80}           & \multicolumn{1}{r|}{4.35}          & \multicolumn{1}{r|}{4.55}          \\ \hline
\multicolumn{1}{|c||}{SURE-IT}        & \multicolumn{1}{r|}{39.27}         & \multicolumn{1}{r|}{49.64}         & \multicolumn{1}{r||}{51.08}         & \multicolumn{1}{r|}{4.57}           & \multicolumn{1}{r|}{7.32}           & \multicolumn{1}{r||}{8.76}           & \multicolumn{1}{r|}{6.82}           & \multicolumn{1}{r|}{13.92}          & \multicolumn{1}{r||}{21.99}          & \multicolumn{1}{r|}{2.38}           & \multicolumn{1}{r|}{2.88}          & \multicolumn{1}{r|}{3.21}          \\ \hline
\multicolumn{1}{|c||}{VDAMP-$\alpha$} & \multicolumn{1}{r|}{\textbf{0.06}} & \multicolumn{1}{r|}{\textbf{0.05}} & \multicolumn{1}{r||}{0.13}          & \multicolumn{1}{r|}{-0.06}          & \multicolumn{1}{r|}{-0.02}          & \multicolumn{1}{r||}{\textbf{-0.03}} & \multicolumn{1}{r|}{\textbf{-0.06}} & \multicolumn{1}{r|}{\textbf{-0.05}} & \multicolumn{1}{r||}{\textbf{-0.09}} & \multicolumn{1}{r|}{0.04}           & \multicolumn{1}{r|}{\textbf{0.00}} & \multicolumn{1}{r|}{\textbf{0.00}} \\ \hline
\multicolumn{1}{|c||}{VDAMP-S}        & \multicolumn{1}{r|}{0.10}          & \multicolumn{1}{r|}{0.07}          & \multicolumn{1}{r||}{\textbf{0.01}} & \multicolumn{1}{r|}{\textbf{-0.03}} & \multicolumn{1}{r|}{\textbf{-0.01}} & \multicolumn{1}{r||}{0.06}           & \multicolumn{1}{r|}{0.20}           & \multicolumn{1}{r|}{0.07}           & \multicolumn{1}{r||}{0.11}           & \multicolumn{1}{r|}{\textbf{0.03}}  & \multicolumn{1}{r|}{0.10}          & \multicolumn{1}{r|}{0.04}          \\ \hline
\multicolumn{1}{l}{}                 & \multicolumn{1}{l}{}                & \multicolumn{1}{l}{}               & \multicolumn{1}{l}{}                & \multicolumn{1}{l}{}               & \multicolumn{1}{l}{}                & \multicolumn{1}{l}{}                & \multicolumn{1}{l}{}                & \multicolumn{1}{l}{}               & \multicolumn{1}{l}{}               & \multicolumn{1}{l}{}                & \multicolumn{1}{l}{}               & \multicolumn{1}{l}{}                \\ \hline
\multicolumn{1}{|c||}{Image}          & \multicolumn{3}{c||}{\textsc{Boat}}                                                                                   & \multicolumn{3}{c||}{\textsc{Cameraman}}                                                                                     & \multicolumn{3}{c||}{\textsc{House}}                                                                                    & \multicolumn{3}{c|}{\textsc{Peppers}}                                                                                   \\ \hline
\multicolumn{1}{|c||}{$n/N$}          & \multicolumn{1}{c|}{4}              & \multicolumn{1}{c|}{6}             & \multicolumn{1}{c||}{8}              & \multicolumn{1}{c|}{4}             & \multicolumn{1}{c|}{6}              & \multicolumn{1}{c||}{8}              & \multicolumn{1}{c|}{4}              & \multicolumn{1}{c|}{6}             & \multicolumn{1}{c||}{8}             & \multicolumn{1}{c|}{4}              & \multicolumn{1}{c|}{6}             & \multicolumn{1}{c|}{8}              \\ \hhline{|=#=|=|=#=|=|=#=|=|=#=|=|=|}
\multicolumn{1}{|c||}{FISTA}          & \multicolumn{1}{r|}{2.17}          & \multicolumn{1}{r|}{3.35}          & \multicolumn{1}{r||}{4.20}          & \multicolumn{1}{r|}{3.70}           & \multicolumn{1}{r|}{5.33}           & \multicolumn{1}{r||}{7.25}           & \multicolumn{1}{r|}{3.01}           & \multicolumn{1}{r|}{5.23}           & \multicolumn{1}{r||}{6.96}           & \multicolumn{1}{r|}{2.90}           & \multicolumn{1}{r|}{5.13}          & \multicolumn{1}{r|}{7.19}          \\ \hline
\multicolumn{1}{|c||}{S-FISTA}        & \multicolumn{1}{r|}{4.79}          & \multicolumn{1}{r|}{5.53}          & \multicolumn{1}{r||}{5.96}          & \multicolumn{1}{r|}{7.26}           & \multicolumn{1}{r|}{8.24}           & \multicolumn{1}{r||}{8.92}           & \multicolumn{1}{r|}{6.14}           & \multicolumn{1}{r|}{7.54}           & \multicolumn{1}{r||}{8.45}           & \multicolumn{1}{r|}{10.04}          & \multicolumn{1}{r|}{11.49}         & \multicolumn{1}{r|}{12.17}         \\ \hline
\multicolumn{1}{|c||}{SURE-IT}        & \multicolumn{1}{r|}{3.30}          & \multicolumn{1}{r|}{4.74}          & \multicolumn{1}{r||}{5.27}          & \multicolumn{1}{r|}{5.68}           & \multicolumn{1}{r|}{8.09}           & \multicolumn{1}{r||}{8.92}           & \multicolumn{1}{r|}{5.26}           & \multicolumn{1}{r|}{7.55}           & \multicolumn{1}{r||}{8.45}           & \multicolumn{1}{r|}{5.21}           & \multicolumn{1}{r|}{10.01}         & \multicolumn{1}{r|}{11.30}         \\ \hline
\multicolumn{1}{|c||}{VDAMP-$\alpha$} & \multicolumn{1}{r|}{0.02}          & \multicolumn{1}{r|}{0.09}          & \multicolumn{1}{r||}{0.05}          & \multicolumn{1}{r|}{\textbf{-0.06}} & \multicolumn{1}{r|}{\textbf{-0.04}} & \multicolumn{1}{r||}{\textbf{0.07}}  & \multicolumn{1}{r|}{\textbf{-0.05}} & \multicolumn{1}{r|}{\textbf{-0.01}} & \multicolumn{1}{r||}{\textbf{-0.01}} & \multicolumn{1}{r|}{\textbf{0.00}}  & \multicolumn{1}{r|}{0.07}          & \multicolumn{1}{r|}{\textbf{0.03}} \\ \hline
\multicolumn{1}{|c||}{VDAMP-S}        & \multicolumn{1}{r|}{\textbf{0.01}} & \multicolumn{1}{r|}{\textbf{0.02}} & \multicolumn{1}{r||}{\textbf{0.01}} & \multicolumn{1}{r|}{-0.08}          & \multicolumn{1}{r|}{0.13}           & \multicolumn{1}{r||}{-0.09}          & \multicolumn{1}{r|}{-0.06}          & \multicolumn{1}{r|}{-0.04}          & \multicolumn{1}{r||}{-0.07}          & \multicolumn{1}{r|}{\textbf{-0.07}} & \multicolumn{1}{r|}{\textbf{0.00}} & \multicolumn{1}{r|}{-0.06}         \\ \hline
\end{tabular}
\caption{\label{tab:mean_kurt_re} Test for Gaussianity using the mean excess kurtosis $\overline{\mathrm{Kurt}}\{\Re[\bm{r}_{k} - \bm{w}_0]\}$ at $K_{it}$. An exact Gaussian has zero mean excess Kurtosis. The smallest absolute values are highlighted in bold.} 
\end{table*}

\begin{figure}[t!]
      \centering
   \subfloat[\label{fig:NMSE_poly}Shepp-Logan, $N/n = 10$]{%
      \includegraphics[width = 0.50 \columnwidth]{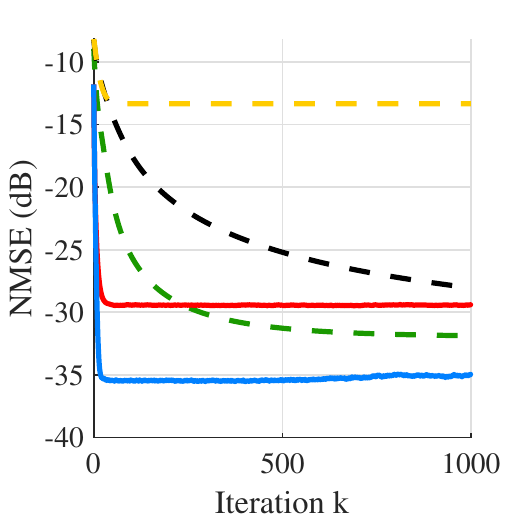}} 
   \subfloat[\label{fig:NMSE_uniform}Cardiac, $N/n = 4$]{%
      \includegraphics[width =0.50 \columnwidth]{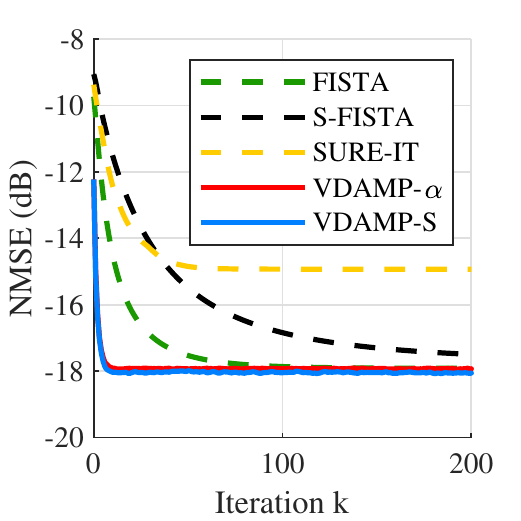}
      \hspace{\fill}}
          \caption{NMSE of FISTA, S-FISTA, SURE-IT, VDAMP-$\alpha$ and VDAMP-S of a Shepp-Logan undersampled at $N/n = 10$ and the cardiac image undersampled at $N/n = 4$. The NMSE at $k=0$ differs between algorithms as the image estimate at $k=0$ is defined to be the after the first thresholding is applied. The cardiac example is shown up to $k=200$, not $K_{it} = 500$, so that the behavior of VDAMP can clearly be seen.}
          \label{fig:NMSE_vs_it}
\end{figure}

\subsection{Time to converge}

Table \ref{tab:conv_time} shows the time to converge for each algorithm, defined as the time taken until the NMSE is within 0.1dB of its value at $K_{it}$. In all test cases, both variations of VDAMP converge considerably more rapidly than the competing FISTA-based methods. Across all experiments, convergence time compared with FISTA was 14.0 times shorter for VDAMP-$\alpha$ and  11.8 times shorter for  VDAMP-S on average, corresponding to a 16.5 and 15.2
 times reduction in the required number of iterations respectively. Of the three FISTA-based algorithms, SURE-IT had the shortest time to convergence, but still required 10.7 and 10.0 times more iterations on average than VDAMP-$\alpha$ and VDAMP-S respectively. Note that the times listed in Table \ref{tab:conv_time} do not include the time required to tune $\lambda$ for FISTA and S-FISTA, nor the time to calculate S-FISTA's subband-weighting $w_b$.

The NMSE vs iteration for the Shepp-Logan undersampled at $N/n = 10$ and the cardiac image undersampled at $N/n = 4$ are shown up to $K_{it}$ and $k=200$ respectively in Fig.  \ref{fig:NMSE_vs_it}, which visualizes VDAMP's comparative rapidity of convergence. In Fig. \ref{fig:cardiac_example}, the cardiac image is shown at $k=10$, where VDAMP-$\alpha$ and VDAMP-S had converged, demonstrating a visible reduction in blocking artifacts for VDAMP.

\subsection{NMSE comparison}

Table \ref{tab:NMSE} shows the normalized mean-squared error (NMSE) $\|\hat{\bm{x}}- \bm{x}_0\|^2_2/\|\bm{x}_0\|^2_2$ of the reconstructed image for each algorithm. The NMSE of FISTA, S-FISTA and VDAMP are generally comparable. Given that VDAMP has 13 model parameters while FISTA has one, one might expect the NMSE of VDAMP would consistently be lower. However, for the Cameraman, the cardiac image at $N/n = 8$, House at $N/n = 4$ and Peppers at $N/n = 4$ FISTA's NMSE was found to be lower. This is due to density compensation in the gradient step, line 4 of Algorithm \ref{alg:VDAMP}, which effectively increases the measurement noise for coefficients sampled with low probability. Note that FISTA's NMSE advantage in these instances may not necessarily arise in realistic, prospectively undersampled reconstruction tasks, as the ground truth cannot be used to hand-tune FISTA's sparse weighting $\lambda$ to a near optimal value, as in the experiments here. Also note that VDAMP-S and VDAMP-$\alpha$ both perform comparatively well on the MRI images, which are of primary importance for the algorithm's intended application.  

Despite employing subband-dependent thesholding, S-FISTA's NMSE at $K_{it}$ was often slightly higher than FISTA. Fig. \ref{fig:wav_err} shows $k=1,2,3$ of VDAMP's wavelet-domain aliasing for the $N/n=12$ Shepp-Logan, which demonstrates that the ratio of the aliasing between subbands may not be constant over iterations. For instance, at $k=1$, the coarse level has greater variance than the fine levels, but at $k=3$ the coarse variance is visibly lower than the fine levels. This is poorly reflected by S-FISTA's threshold weighting, which is fixed over iterations and not dependent on the current estimate. Further, \cite{Bayram2010a} notes that while \eqref{eqn:s-fista_weight} is sufficient to ensure convergence, the inequality is not tight so may lead to weights that are smaller than necessary, which slows convergence. In \cite{Guerquin-Kern2011}, which uses different sampling schemes to that employed here, it was found that S-FISTA performed slightly better than FISTA, suggesting that S-FISTA's relative performance may be particularly dependent on the sampling scheme employed. The comparatively poor performance of SURE-IT highlights the need for zero-mean Gaussian aliasing for effective automatic parameter tuning with SURE.

\subsection{Empirical evidence of state evolution \label{sec:emp_evidence}}

This section presents empirical evidence that VDAMP obeys the colored state evolution given by \eqref{eqn:VDAMPse} using kurtosis and quantile-quantile plots. The excess kurtosis of a real random variable $X$ is defined as $\mathrm{Kurt}\{X\} = \mu_4/\sigma^4 - 3$, where $\mu_4$ is the fourth central moment and $\sigma$ is the standard deviation. The Gaussianity of the aliasing of $\bm{r}_k$ was tested by calculating the mean of per-subband empirical kurtosis of the real part,
\begin{equation*}
	\overline{\mathrm{Kurt}}\{\Re[\bm{r}_{k} - \bm{w}_0]\} = \frac{1}{3s+1} \sum_{b=1}^{3s+1} \mathrm{Kurt}\{ \Re[\bm{r}_{k, b} - \bm{w}_{0,b}]\},
\end{equation*}
and comparing to zero, which is the kurtosis of a white Gaussian distribution. 
Table \ref{tab:mean_kurt_re} shows the mean kurtosis for all images and sampling factors at $k = K_{it}$. The proximity to zero is consistent with a colored state evolution for all image types and undersampling factors, and for both VDAMP-$\alpha$ and VDAMP-S, confirming that the difference in performance between the algorithms is not due to a breakdown in state evolution. The imaginary part, which is not included here for conciseness, was found to have a similarly small mean excess kurtosis. 

\begin{figure}[b!]
\captionsetup[subfigure]{labelformat=empty}
\hspace{\fill}
   \subfloat[\label{fig:r1}(a) $|\bm{r}_1 - \bm{w}_0|$]{%
      \includegraphics[height = 0.27 \columnwidth]{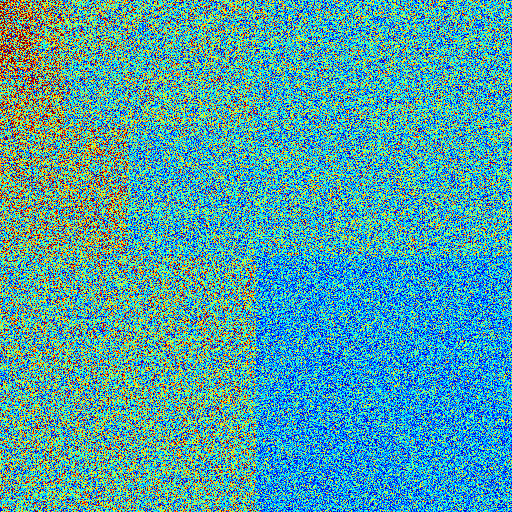}}
\hspace{\fill}
   \subfloat[\label{fig:r2}(b) $|\bm{r}_2 - \bm{w}_0|$]{%
      \includegraphics[height=0.27 \columnwidth]{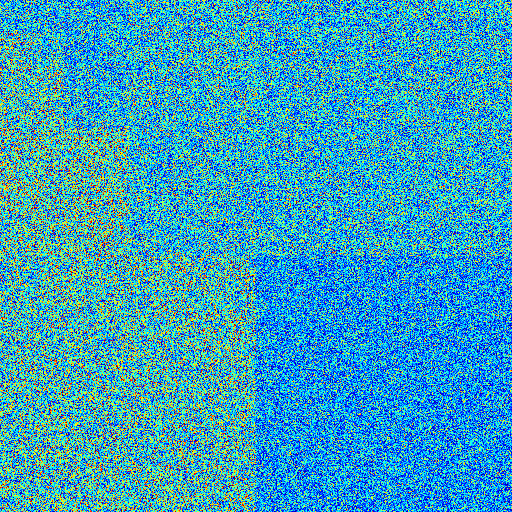}}
\hspace{\fill} 
   \subfloat[\label{fig:r3}(c) $|\bm{r}_3 - \bm{w}_0|$]{%
      \includegraphics[height=0.27 \columnwidth]{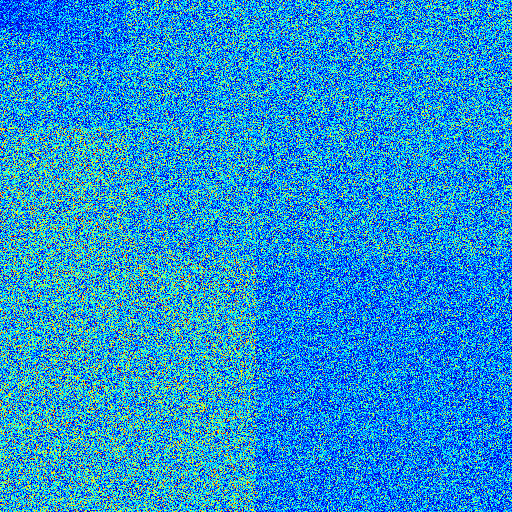}}
      \hspace{0.1cm} 
   \subfloat[]{%
      \includegraphics[height= 0.28 \columnwidth]{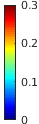} }
      \hspace{\fill} 
      \caption{\label{fig:wav_err} The magnitude of the effective noise for VDAMP-S for a Shepp-Logan undersampled with $N/n = 12$ for iterations $k = 1, 2, 3$, where the colorbar shows the proportion of the maximum of $\bm{x}_0$.}
\end{figure}

\begin{figure}
\centering
    \includegraphics[width=1\columnwidth]{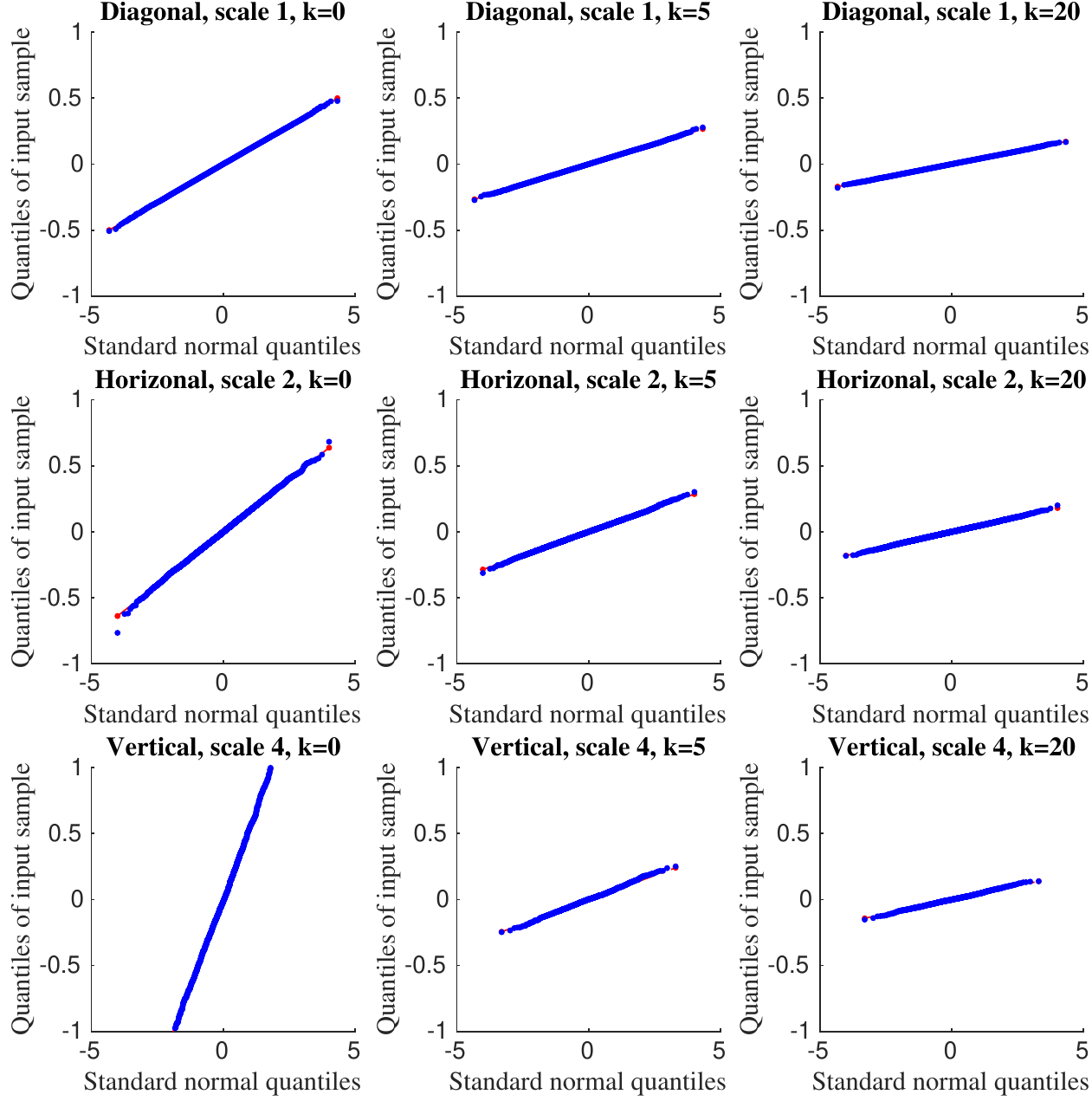}
    \caption{Normalized quantile-quantile plots against a Gaussian for three subbands of VDAMP-S's $\bm{r}_{k}-\bm{w}_0$ at $k= 0,5,20$ for the Shepp-Logan sampled with $N/n = 12$ in blue, and points along a straight line in red. The real part is plotted in the top and bottom rows and the imaginary is plotted in the middle row. Linearity of the blue points indicates that that the data comes from a Gaussian distribution, and the decreasing gradient shows that the variance decreases with increasing $k$. Finite dimensional effects causing small deviations from an exact Gaussian are more apparent at coarse scales, where the dimension is smaller.}
    \label{fig:qqplots}
\end{figure}

Using the example of the Shepp-Logan undersampled with $N/n = 12$, Fig. \ref{fig:wav_err} shows VDAMP-S's $|\bm{r}_{k} - \bm{w}_0|$ for $k=1, 2, 3$, visualizing the preservation of the unbiased subband-dependent aliasing structure shown for uniform sampling in Fig. \ref{fig:intuition}. For $k = 0, 5, 20$, Fig. \ref{fig:qqplots} shows quantile-quantile plots against a Gaussian of the three illustrative subbands of $\bm{r}_{k} - \bm{w}_0$: the diagonal detail at scale 1, the horizontal detail at scale 2 and the vertical detail at scale 4, where scale 1 is the finest and scale 4 is the coarsest. The linearity of the blue points provides strong evidence that the per-subband effective noise is Gaussian.  

\begin{figure*}[h!]
\centering
    \includegraphics[width=1.3\columnwidth]{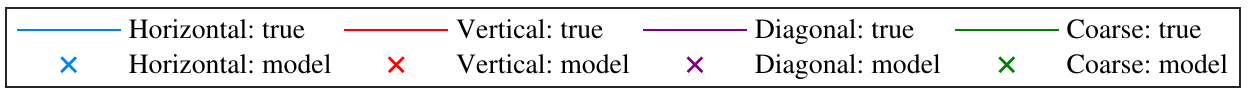}
    \includegraphics[width=2\columnwidth]{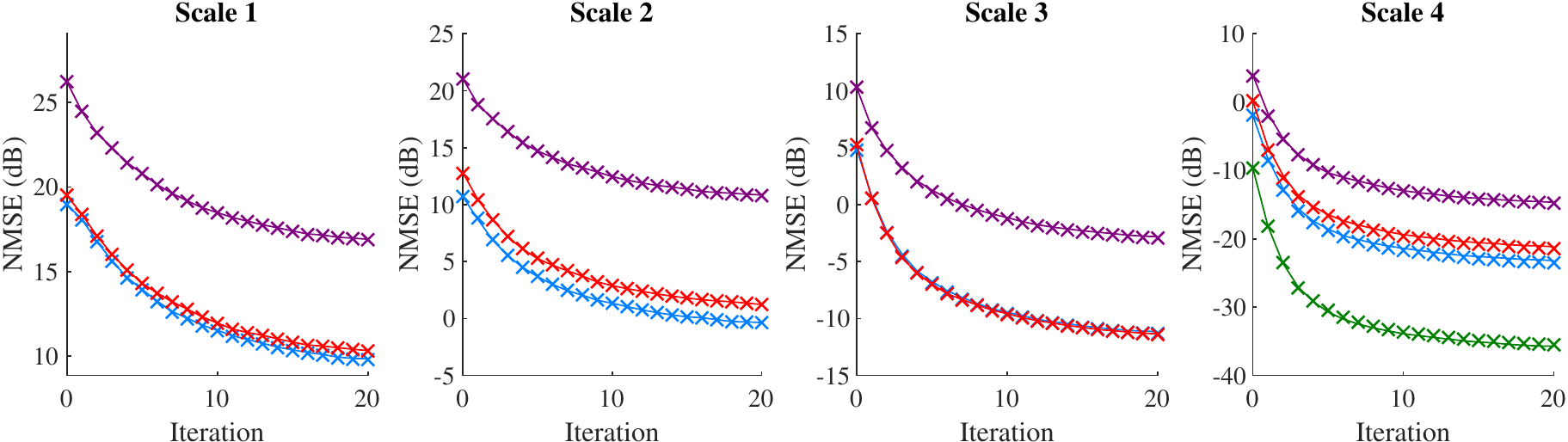}
    \caption{Per subband NMSE $\|\bm{r}_{k,b} - \bm{w}_{0,b}\|^2_2/\|\bm{w}_{0,b}\|^2_2$ versus iteration index $k$ for a $N/n = 12$  undersampled Shepp-Logan reconstructed with VDAMP-S. Lines show the actual NMSE and crosses show the predictions from $\bm{\tau}_{k}$.}
    \label{fig:nmse_tracking}
\end{figure*}

The efficacy of automatic threshold selection with cSURE depends on how accurately the diagonal of $\bm{\Sigma}_k^2$ is modeled by $\bm{\tau}_k$. For $k = 0, 1, \ldots, 20$, Fig. \ref{fig:nmse_tracking} shows the ground truth subband NMSE $\|\bm{r}_{k,b} - \bm{w}_{0,b}\|^2_2/\|\bm{w}_{0,b}\|^2_2$  at all four scales and the prediction of VDAMP, where the NMSE is per subband. The true NMSE is closely tracked by $\bm{\tau}_k$ at all scales, which implies that parameter selection with SURE is truly near-optimal. Since, by Appendix \ref{app:tau_transform}, $\bm{\tau}_k$ is unbiased when the aliasing of $\bm{r}_k$ is zero mean and i.i.d., Fig. \ref{fig:nmse_tracking} also provides further verification of colored state evolution.

\section{Conclusions}
Based on the observation that Fourier sampling from a non-uniform spectral density leads to colored aliasing, we propose VDAMP, an algorithm based on OAMP that obeys a colored state evolution.  State evolution provides an effective way to tune model parameters via cSURE, implying that a single algorithm can be used for arbitrary variable density scheme and image type without the need for manual adjustment.  More degrees of freedom are feasibly allowed in the model, enabling higher order prior information such as anisotropy, variability across scales and structured sparsity, without the need to estimate the structure a priori \cite{Baraniuk2010}. 

VDAMP was motivated by the application of compressed sensing to accelerated MRI. Developments are required for VDAMP to be applicable to MRI data acquired across multiple coils that possess a sensitivity profile\cite{Liang2009, Pruessmann2001}, and for VDAMP to be applicable to Fourier sampling with 1D readout curves, where elements of $\Omega$ are generated dependently.

It is known that the state evolution of OAMP holds for a wide range of denoisers $\bm{g}(\bm{r}_{k}; \tau_{k})$ \cite{Metzler2016, Xue2016}. In \cite{Ahmad2020}, a number of standard compressed sensing algorithms that leverage image denoisers designed for Gaussian noise were shown to perform well on MRI reconstruction tasks, despite the mismatch between the aliasing and its model. A sophisticated denoiser equipped to deal with wavelet coefficients corrupted with known colored Gaussian noise would be expected to perform well in conjunction with VDAMP. There has also been interest in algorithms that leverage the statistical modeling capabilities of neural networks  \cite{LISTA2010, NIPS2017_6774, Ito2019, Ahmad2020}. VDAMP with a neural network denoiser $\bm{g}(\bm{r}_{k}; \bm{\tau}_{k})$ could accommodate ground-truth free training by using cSURE as the loss, as shown for AMP in \cite{Metzler2018, Zhussip2019}.

\begin{appendices}
\section{The power spectrum of the aliasing of \texorpdfstring{$\widetilde{\bm{x}}$}{x_tilde} \label{app:power_spec}}

This appendix proves that the magnitude of the entry-wise difference between $\bm{y}_0$ and the unbiased estimate $\bm{P}^{-1} \bm{y}$ is
\begin{align}
    \mathds{E}_{\Omega, \varepsilon} \{ |\bm{y}_0 - \bm{P}^{-1}\bm{y}|^2\}  = ( \bm{P}^{-1} - \mathds{1}_N)|\bm{y}_0|^2 + \sigma^2_{\varepsilon}   \bm{P}^{-1} \bm{1}_N \label{eqn:app_pow1}
\end{align}    
\textit{Proof:} Since the entries of $\bm{y}$ are independent and $\bm{P}^{-1}$ is diagonal, we can consider \eqref{eqn:app_pow1} as $N$ distinct one-dimensional expressions. The $i$th entry of  \eqref{eqn:app_pow1} is
\begin{align*}
     &\mathds{E}_{m_i, \varepsilon_i} \left\{ \left|y_{0,i} - \frac{y_i}{p_i} \right|^2\right\} \\ 
     =&  \mathds{E}_{m_i, \varepsilon_i} \left\{ \left|y_{0,i} - \frac{m_i}{p_i}(y_{0,i} + \varepsilon_i) \right|^2 \right\} \\
    =& \mathds{E}_{m_i, \varepsilon_i} \left\{ \left|\left(1- \frac{m_i}{p_i}\right)y_{0,i} - \frac{m_i}{p_i}\varepsilon_i\right|^2\right\}, \stepcounter{equation}\tag{\theequation}\label{eqn:app_pow2}
\end{align*}
where $m_i$ is the $i$th diagonal of $\bm{M}_\Omega$. By assumption, $m_i$ is distributed according to a Bernoulli distribution with $\mathds{E}_{m_i}\{ m_i\} = p_i$. The expectation over $m_i$ can therefore be found by resolving  \eqref{eqn:app_pow2} at $m_i =1$ and $m_i = 0$ and summing with weights $p_i$ and $1-p_i$ respectively:
\begin{multline*}
    \mathds{E}_{m_i, \varepsilon_i} \left\{ \left|\left(1- \frac{m_i}{p_i}\right)y_{0,i} - \frac{m_i}{p_i}\varepsilon_i \right|^2 \right\} \\ =\mathds{E}_{\varepsilon_i} \left\{  p_i\left|\left(\frac{1-p_i}{p_i}\right)y_{0,i} - \frac{\varepsilon_i}{p_i}\right|^2 + (1-p_i) |y_{0,i}|^2\right\}.
\end{multline*}
Expanding the first term, and noting that $\mathds{E}_{\varepsilon_i} \{\varepsilon_i\} = 0$ and $\mathds{E}_{\varepsilon_i}\{|\varepsilon_i|^2\} = \sigma_\varepsilon^2$, 
\begin{align*}
    & \mathds{E}_{\varepsilon_i} \left \{  p_i\left |\left(\frac{1-p_i}{p_i}\right)y_{0,i} - \frac{\varepsilon_i}{p_i}\right|^2 + (1-p_i) |y_{0,i}|^2 \right\} 
    \\ =&  \begin{multlined}[t] \mathds{E}_{\varepsilon_i} \biggl\{  p_i\left|\left(\frac{1-p_i}{p_i}\right)y_{0,i}\right|^2 + p_i\left|\frac{\varepsilon_i}{p_i}  \right|^2 
    \\ - \left(\frac{1-p_i}{p_i}\right)(y_{0,i}^* \varepsilon_i + y_{0,i} \varepsilon_i^*) + (1-p_i) |y_{0,i}|^2 \biggr\}
    \end{multlined}
    \\ =&  \frac{(1-p_i)^2}{p_i}|y_{0,i}|^2 + \frac{\sigma_\varepsilon^2}{p_i} + (1-p_i) |y_{0,i}|^2 \\
    =& \left(\frac{1-p_i}{p_i}\right)|y_{0,i}|^2 + \frac{\sigma_\varepsilon^2}{p_i}
\end{align*}
which is the $i$th entry of the right-hand-side of \eqref{eqn:app_pow1}. This completes the proof.

\section{Proof that \texorpdfstring{$\bm{\tau}_k^y$}{tau} is unbiased \label{app:unbiased_tau_update} }

This appendix shows that the $\bm{\tau}_k^y$ update, \eqref{eqn:tauk}, is an unbiased estimate of the power spectrum of the aliasing of $\bm{r}_k$, \eqref{eqn:exp_y}:
\begin{multline}
\mathds{E} \{ \bm{M}_\Omega \bm{P}^{-1}[(\bm{P}^{-1} - \mathds{1}_N) |\bm{z}_k|^2 + \sigma_\varepsilon^2 \bm{1}_N]\} \\ = (\bm{P}^{-1} - \mathds{1}_N) |\bm{F}\bm{\Psi}^H \widetilde{\bm{r}}_k - \bm{y}_0 |^2 + \sigma_\varepsilon^2 \bm{P}^{-1}\bm{1}_N \label{eqn:app_tau_main}
\end{multline}

\textit{Proof:} Defining $\widetilde{\bm{y}}_k = \bm{F}\bm{\Psi}^H \widetilde{\bm{r}}_k$, the $i$th entry of the left-hand-side of  \eqref{eqn:app_tau_main} in terms of $\bm{y}_0$ is 
\begin{multline*}
\mathds{E}_{m_i, \varepsilon_i} \left\{ \frac{m_i}{p_i}\left[ \left(\frac{1-p_i}{p_i}\right)|y_i - m_i\widetilde{y}_{k,i}|^2 + \sigma_\varepsilon^2\right] \right\} \\
= \mathds{E}_{m_i, \varepsilon_i} \left\{ \frac{m_i}{p_i}\left[ \left(\frac{1-p_i}{p_i}\right)|m_i(y_{0,i} + \varepsilon_i) - m_i\widetilde{y}_{k,i}|^2 + \sigma_\varepsilon^2\right] \right\}
\end{multline*}
The expectation over $m_i$ is 

\begin{multline*}
 \mathds{E}_{m_i, \varepsilon_i} \left\{ \frac{m_i}{p_i}\left[ \left(\frac{1-p_i}{p_i}\right)|m_i(y_{0,i} + \varepsilon_i) - m_i\widetilde{y}_{k,i}|^2 + \sigma_\varepsilon^2\right] \right\} \\
=  \mathds{E}_{\varepsilon_i} \left\{ \left(\frac{1-p_i}{p_i}\right)|y_{0,i} + \varepsilon_i - \widetilde{y}_{k,i}|^2 +\sigma_\varepsilon^2 \right\}
\end{multline*}
Since $\mathds{E}_{\varepsilon_i} \{\varepsilon_i\} = 0$ and $\mathds{E}_{\varepsilon_i}\{|\varepsilon_i|^2\} = \sigma_\varepsilon^2$,  
 \begin{align*} 
& \mathds{E}_{\varepsilon_i} \left\{ \left(\frac{1-p_i}{p_i}\right)|y_{0,i} + \varepsilon_i - \widetilde{y}_{k,i}|^2 +\sigma_\varepsilon^2 \right\} \\
&=  \left(\frac{1-p_i}{p_i}\right)(|y_{0,i} - \widetilde{y}_{k,i}|^2 + \sigma_\varepsilon^2) +\sigma_\varepsilon^2 \\
& = \left(\frac{1-p_i}{p_i}\right)|y_{0,i} - \widetilde{y}_{k,i}|^2 + \frac{\sigma_\varepsilon^2}{p_i}
\end{align*}
which is the $i$th entry of the right-hand-side of \eqref{eqn:app_tau_main}. This completes the proof. 

\section{Transforming \texorpdfstring{$\bm{\tau}_k^y$}{tau} to the wavelet domain \label{app:tau_transform}}

This appendix proves that when $\mathds{E} \{\bm{r}_k\}=\bm{w}_0$, and the entries of $\bm{r}_k$ are independent,
\begin{equation}
	\mathds{E}\{|\bm{r}_k - \bm{w}_0|^2\} = |\bm{\Psi F}^H|^2 \mathds{E} \{\bm{\tau}_k^y\}
\end{equation}
where $\bm{\tau}_k^y$ is defined in \eqref{eqn:tauk}. 

\textit{Proof:} 
Let the wavelet-domain error be $\bm{r}_k - \bm{w}_0 = \bm{A u}$, where $\bm{u} = \bm{F \Psi}^H \bm{r}_k - \bm{y}_0$ is the Fourier-domain residual, and $\bm{A} =  \bm{ \Psi F}^H$, where the iteration index $k$ has been removed to simplify notation. The $i$th entry of $|\bm{Au}|^2$ is
\begin{align*}
    |\sum_j A_{ij} u_j|^2 &= (\sum_j  A_{ij} u_j) (\sum_{l}  A_{il}^* u_l^*)\\
    &= \sum_j (A_{ij}A_{ij}^* u_j u_j^* + \sum_{l \neq j} A_{ij}A_{il}^* u_j u_l^*)
\end{align*}

Since, by assumption, $\bm{r}_k$ is unbiased and independent, the $l\neq j$ terms are zero in expectation. Therefore
\begin{align*}
    \mathds{E}\{|\sum_j A_{ij} u_j|^2\} & =   \mathds{E}\{ \sum_j A_{ij}A_{ij}^* u_j u_j^*\} \\
    &=      \sum_j |A_{ij}|^2 \mathds{E}\{|u_j|^2 \}. 
\end{align*}
Since, by Appendix \ref{app:unbiased_tau_update}, $\mathds{E}\{|\bm{u}|^2 \} = \mathds{E}\{\bm{\tau}_k^y \}$, we have
\begin{align*}
    \mathds{E}\{|\bm{A u}|^2\} &=  |\bm{A}|^2 \mathds{E}\{|\bm{u}|^2\} = |\bm{\Psi F}^H|^2 \mathds{E}\{\bm{\tau}_k^y\}
\end{align*}
This completes the proof.

\section{SURE for complex variables \label{app:cSURE}}
This appendix proves that cSURE, defined in \eqref{eqn:cSURE_general}, is an unbiased estimate of the risk, so that 
\begin{equation}
 \mathds{E}\{\| \bm{d}(\bm{v}) - \bm{v}_0\|^2_2\} = \mathds{E}\{cSURE(\bm{d}(\bm{v}))\}
\end{equation}
where the expectation is over $\bm{v} = \bm{v}_0 + \mathcal{CN}(\bm{0}, \tau_v \mathds{1}_{N_v})$ and $\bm{d}(\bm{v}) = \bm{v} + \bm{h}(\bm{v})$ is a denoiser. The proof in this appendix is a complex noise variation on the standard proof of SURE, as found in \cite{Stein1981, Luisier2007}.

\textit{Proof:} 
The Euclidean distance between the ground truth $\bm{v}_0$ and the denoised vector $\bm{d}(\bm{v})$ can be expanded as
\begin{multline}
	\mathds{E}\{\| \bm{d}(\bm{v}) - \bm{v}_0\|^2_2\} = \mathds{E}\{\|\bm{h}(\bm{v})\|^2_2 + \|\bm{v} -\bm{v}_0\|^2_2 \\ - 2( \Re[\bm{h}(\bm{v})]^H \Re[\bm{v} - \bm{v}_0] + \Im [\bm{h}(\bm{v})]^H \Im[\bm{v} - \bm{v}_0])\} \label{eqn:cSURE_expanded}
\end{multline}
By the noise model of $\bm{v}$, the second term on the right hand side is 
\begin{equation}
\mathds{E}\{\|\bm{v} -\bm{v}_0\|^2_2\} = N_v \tau_v
\end{equation}
By Stein's lemma \cite{Stein1981} (see also (6) of \cite{Luisier2007}),  and recalling that $\mathcal{CN}(\bm{0}, \tau_v \mathds{1}_{N_v})$ is defined such that the variance of the real and imaginary parts is $\tau_v/2$, the final term of the right-hand-side of \eqref{eqn:cSURE_expanded} is
\begin{align*}
	\mathds{E}\{\Re[\bm{h}(\bm{v})]^H \Re[\bm{v} - \bm{v}_0]\} & = \frac{\tau_v}{2} \mathds{E}\left\{\sum_j \frac{\partial \Re[h_j(\bm{v})]}{\partial \Re [v_j]}\right\} \\
	 & =  \frac{\tau_v}{2}\left(N_v -\mathds{E}\left\{ \sum_j \frac{\partial \Re[d_j(\bm{v})]}{\partial \Re [v_j]} \right\} \right)
\end{align*}
and similarly for the imaginary part. Overall, \eqref{eqn:cSURE_expanded} is therefore
\begin{align*}
	&\mathds{E}\{\| \bm{d}(\bm{v}) - \bm{v}_0\|^2_2\} \\ =	&\mathds{E} \Bigg\{\|\bm{h}(\bm{v})\|^2_2 - N_v \tau_v 
		 + \tau_v \sum_j \left(\frac{\partial \Re[d_j(\bm{v})]}{\partial \Re [v_j]} + \frac{\partial \Im[d_j(\bm{v})]}{\partial \Im [v_j]}\right) \Bigg\} \\
		 =&\mathds{E}\{ \|\bm{h}(\bm{v})\|^2_2  + N_v \tau_v [2\braket{\bm{\partial} (\bm{d}(\bm{v}))} - 1] \}
\end{align*}
which is the expectation of cSURE, given in \eqref{eqn:cSURE_general}. This completes the proof.

\end{appendices}

\bibliographystyle{IEEEtran}
\bibliography{library_fixed}

\begin{thebibliography}{10}
\providecommand{\url}[1]{#1}
\csname url@samestyle\endcsname
\providecommand{\newblock}{\relax}
\providecommand{\bibinfo}[2]{#2}
\providecommand{\BIBentrySTDinterwordspacing}{\spaceskip=0pt\relax}
\providecommand{\BIBentryALTinterwordstretchfactor}{4}
\providecommand{\BIBentryALTinterwordspacing}{\spaceskip=\fontdimen2\font plus
\BIBentryALTinterwordstretchfactor\fontdimen3\font minus
  \fontdimen4\font\relax}
\providecommand{\BIBforeignlanguage}[2]{{%
\expandafter\ifx\csname l@#1\endcsname\relax
\typeout{** WARNING: IEEEtran.bst: No hyphenation pattern has been}%
\typeout{** loaded for the language `#1'. Using the pattern for}%
\typeout{** the default language instead.}%
\else
\language=\csname l@#1\endcsname
\fi
#2}}
\providecommand{\BIBdecl}{\relax}
\BIBdecl

\bibitem{Millard2019}
\BIBentryALTinterwordspacing
C.~Millard, A.~T. Hess, B.~Mailh{\'{e}}, and J.~Tanner, ``{An Approximate
  Message Passing Algorithm for Rapid Parameter-Free Compressed Sensing MRI},''
  no.~4, 2019. [Online]. Available: \url{http://arxiv.org/abs/1911.01234}
\BIBentrySTDinterwordspacing

\bibitem{Donoho2006}
\BIBentryALTinterwordspacing
D.~L. Donoho, ``{Compressed sensing},'' \emph{IEEE Transactions on Information
  Theory}, vol.~52, no.~4, pp. 1289--1306, apr 2006. [Online]. Available:
  \url{http://ieeexplore.ieee.org/document/1614066/}
\BIBentrySTDinterwordspacing

\bibitem{Candes2006}
\BIBentryALTinterwordspacing
E.~Candes, J.~Romberg, and T.~Tao, ``{Robust uncertainty principles: exact
  signal reconstruction from highly incomplete frequency information},''
  \emph{IEEE Transactions on Information Theory}, vol.~52, no.~2, pp. 489--509,
  feb 2006. [Online]. Available:
  \url{http://ieeexplore.ieee.org/document/1580791/}
\BIBentrySTDinterwordspacing

\bibitem{Lustig2007}
\BIBentryALTinterwordspacing
M.~Lustig, D.~Donoho, and J.~M. Pauly, ``{Sparse MRI: The application of
  compressed sensing for rapid MR imaging},'' \emph{Magnetic Resonance in
  Medicine}, vol.~58, no.~6, pp. 1182--1195, dec 2007. [Online]. Available:
  \url{http://doi.wiley.com/10.1002/mrm.21391}
\BIBentrySTDinterwordspacing

\bibitem{Otazo2010}
\BIBentryALTinterwordspacing
R.~Otazo, D.~Kim, L.~Axel, and D.~K. Sodickson, ``{Combination of compressed
  sensing and parallel imaging for highly accelerated first-pass cardiac
  perfusion MRI},'' \emph{Magnetic Resonance in Medicine}, vol.~64, no.~3, pp.
  767--776, sep 2010. [Online]. Available:
  \url{http://www.ncbi.nlm.nih.gov/pubmed/20535813}
\BIBentrySTDinterwordspacing

\bibitem{Jaspan2015}
\BIBentryALTinterwordspacing
O.~N. Jaspan, R.~Fleysher, and M.~L. Lipton, ``{Compressed sensing MRI: a
  review of the clinical literature},'' \emph{The British Journal of
  Radiology}, vol.~88, no. 1056, p. 20150487, dec 2015. [Online]. Available:
  \url{http://www.birpublications.org/doi/10.1259/bjr.20150487}
\BIBentrySTDinterwordspacing

\bibitem{Ye2019}
\BIBentryALTinterwordspacing
J.~C. Ye, ``{Compressed sensing MRI: a review from signal processing
  perspective},'' \emph{BMC Biomedical Engineering}, vol.~1, no.~1, p.~8, dec
  2019. [Online]. Available:
  \url{https://bmcbiomedeng.biomedcentral.com/articles/10.1186/s42490-019-0006-z}
\BIBentrySTDinterwordspacing

\bibitem{Donoho2017}
\BIBentryALTinterwordspacing
D.~Donoho, ``{How High-Dimensional Geometry is Transforming the MRI
  Industry},'' 2017. [Online]. Available: \url{https://vimeo.com/225634059}
\BIBentrySTDinterwordspacing

\bibitem{Puy2011}
\BIBentryALTinterwordspacing
G.~Puy, P.~Vandergheynst, and Y.~Wiaux, ``{On Variable Density Compressive
  Sampling},'' \emph{IEEE Signal Processing Letters}, vol.~18, no.~10, pp.
  595--598, oct 2011. [Online]. Available:
  \url{http://ieeexplore.ieee.org/document/5976374/}
\BIBentrySTDinterwordspacing

\bibitem{wangVDS}
Z.~Wang and G.~R. Arce, ``{Variable Density Compressed Image Sampling},''
  \emph{IEEE Transactions on Image Processing}, vol.~19, no.~1, pp. 264--270,
  jan 2010.

\bibitem{Krahmer2014}
F.~Krahmer and R.~Ward, ``{Stable and robust sampling strategies for
  compressive imaging},'' \emph{IEEE Transactions on Image Processing},
  vol.~23, no.~2, pp. 612--622, 2014.

\bibitem{Chauffert2013}
N.~Chauffert, P.~Ciuciu, and P.~Weiss, ``{Variable density compressed sensing
  in MRI. Theoretical vs heuristic sampling strategies},'' \emph{Proceedings -
  International Symposium on Biomedical Imaging}, pp. 298--301, 2013.

\bibitem{Adcock2017}
\BIBentryALTinterwordspacing
B.~Adcock, A.~C. Hansen, C.~Poon, and B.~Roman, ``{Breaking the coherence
  barrier: a new theory for compressed sensing},'' \emph{Forum of Mathematics,
  Sigma}, vol.~5, p.~e4, feb 2017. [Online]. Available:
  \url{https://www.cambridge.org/core/product/identifier/S2050509416000323/type/journal\_article}
\BIBentrySTDinterwordspacing

\bibitem{Donoho2009}
\BIBentryALTinterwordspacing
D.~L. Donoho, A.~Maleki, and A.~Montanari, ``{Message Passing Algorithms for
  Compressed Sensing},'' jul 2009. [Online]. Available:
  \url{http://dx.doi.org/10.1073/pnas.0909892106}
\BIBentrySTDinterwordspacing

\bibitem{Bayati2011}
\BIBentryALTinterwordspacing
M.~Bayati and A.~Montanari, ``{The Dynamics of Message Passing on Dense Graphs,
  with Applications to Compressed Sensing},'' \emph{IEEE Transactions on
  Information Theory}, vol.~57, no.~2, pp. 764--785, feb 2011. [Online].
  Available: \url{http://ieeexplore.ieee.org/document/5695122/}
\BIBentrySTDinterwordspacing

\bibitem{Bayati2015}
\BIBentryALTinterwordspacing
M.~Bayati, M.~Lelarge, and A.~Montanari, ``{Universality in polytope phase
  transitions and message passing algorithms},'' \emph{The Annals of Applied
  Probability}, vol.~25, no.~2, pp. 753--822, apr 2015. [Online]. Available:
  \url{http://projecteuclid.org/euclid.aoap/1424355130}
\BIBentrySTDinterwordspacing

\bibitem{Rangan2014}
\BIBentryALTinterwordspacing
S.~Rangan, P.~Schniter, and A.~Fletcher, ``{On the convergence of approximate
  message passing with arbitrary matrices},'' in \emph{2014 IEEE International
  Symposium on Information Theory}.\hskip 1em plus 0.5em minus 0.4em\relax
  IEEE, jun 2014, pp. 236--240. [Online]. Available:
  \url{http://ieeexplore.ieee.org/lpdocs/epic03/wrapper.htm?arnumber=6874830}
\BIBentrySTDinterwordspacing

\bibitem{Caltagirone2014}
\BIBentryALTinterwordspacing
F.~Caltagirone, L.~Zdeborova, and F.~Krzakala, ``{On convergence of approximate
  message passing},'' in \emph{2014 IEEE International Symposium on Information
  Theory}.\hskip 1em plus 0.5em minus 0.4em\relax IEEE, jun 2014, pp.
  1812--1816. [Online]. Available:
  \url{http://ieeexplore.ieee.org/lpdocs/epic03/wrapper.htm?arnumber=6875146}
\BIBentrySTDinterwordspacing

\bibitem{Guo2015}
\BIBentryALTinterwordspacing
C.~Guo and M.~E. Davies, ``{Near Optimal Compressed Sensing Without Priors:
  Parametric SURE Approximate Message Passing},'' \emph{IEEE Transactions on
  Signal Processing}, vol.~63, no.~8, pp. 2130--2141, apr 2015. [Online].
  Available: \url{http://ieeexplore.ieee.org/document/7054509/}
\BIBentrySTDinterwordspacing

\bibitem{Rangan2016}
\BIBentryALTinterwordspacing
S.~Rangan, P.~Schniter, E.~Riegler, A.~K. Fletcher, and V.~Cevher, ``{Fixed
  Points of Generalized Approximate Message Passing With Arbitrary Matrices},''
  \emph{IEEE Transactions on Information Theory}, vol.~62, no.~12, pp.
  7464--7474, dec 2016. [Online]. Available:
  \url{http://ieeexplore.ieee.org/document/7600404/}
\BIBentrySTDinterwordspacing

\bibitem{Ma2017}
\BIBentryALTinterwordspacing
J.~Ma and L.~Ping, ``{Orthogonal AMP},'' \emph{IEEE Access}, vol.~5, pp.
  2020--2033, 2017. [Online]. Available:
  \url{http://ieeexplore.ieee.org/document/7817805/}
\BIBentrySTDinterwordspacing

\bibitem{Rangan2019}
\BIBentryALTinterwordspacing
S.~Rangan, P.~Schniter, and A.~K. Fletcher, ``{Vector Approximate Message
  Passing},'' \emph{IEEE Transactions on Information Theory}, pp. 1--1, 2019.
  [Online]. Available: \url{https://ieeexplore.ieee.org/document/8713501/}
\BIBentrySTDinterwordspacing

\bibitem{Daubechies2004}
\BIBentryALTinterwordspacing
I.~Daubechies, M.~Defrise, and C.~{De Mol}, ``{An iterative thresholding
  algorithm for linear inverse problems with a sparsity constraint},''
  \emph{Communications on Pure and Applied Mathematics}, vol.~57, no.~11, pp.
  1413--1457, nov 2004. [Online]. Available:
  \url{http://doi.wiley.com/10.1002/cpa.20042}
\BIBentrySTDinterwordspacing

\bibitem{Stein1981}
\BIBentryALTinterwordspacing
C.~M. Stein, ``{Estimation of the Mean of a Multivariate Normal
  Distribution},'' \emph{The Annals of Statistics}, vol.~9, no.~6, pp.
  1135--1151, nov 1981. [Online]. Available:
  \url{http://projecteuclid.org/euclid.aos/1176345632}
\BIBentrySTDinterwordspacing

\bibitem{Xue2016}
\BIBentryALTinterwordspacing
Z.~Xue, J.~Ma, and X.~Yuan, ``{D-OAMP: A denoising-based signal recovery
  algorithm for compressed sensing},'' in \emph{2016 IEEE Global Conference on
  Signal and Information Processing (GlobalSIP)}.\hskip 1em plus 0.5em minus
  0.4em\relax IEEE, dec 2016, pp. 267--271. [Online]. Available:
  \url{http://ieeexplore.ieee.org/document/7905845/}
\BIBentrySTDinterwordspacing

\bibitem{Goossens2011}
B.~Goossens, J.~Aelterman, H.~Luong, A.~Pizurica, and W.~Philips,
  ``{Wavelet-Based Analysis and Estimation of Colored Noise},'' in
  \emph{Discrete Wavelet Transforms - Algorithms and Applications},
  H.~Olkkonen, Ed.\hskip 1em plus 0.5em minus 0.4em\relax Intech, 2011, ch.~15,
  pp. 255--280.

\bibitem{Li2010}
T.~Li, M.~Wang, and W.~Xiong, ``{A novel method for filtering of Gaussian
  colored noise in images with wavelet transform},'' \emph{ICEIE 2010 - 2010
  International Conference on Electronics and Information Engineering,
  Proceedings}, vol.~1, no. Iceie, pp. 184--189, 2010.

\bibitem{Johnstone1997}
I.~M. Johnstone and B.~W. Silverman, ``{Wavelet threshold estimators for data
  with correlated noise},'' \emph{Journal of the Royal Statistical Society.
  Series B: Statistical Methodology}, vol.~59, no.~2, pp. 319--351, 1997.

\bibitem{Metzler2016}
C.~A. Metzler, A.~Maleki, and R.~G. Baraniuk, ``{From Denoising to Compressed
  Sensing},'' \emph{IEEE Transactions on Information Theory}, vol.~62, no.~9,
  pp. 5117--5144, 2016.

\bibitem{Schniter2017a}
\BIBentryALTinterwordspacing
P.~Schniter, S.~Rangan, and A.~Fletcher, ``{Plug-and-play Image Recovery using
  Vector AMP}.''\hskip 1em plus 0.5em minus 0.4em\relax BASP Frontiers Workshop
  2017, 2017. [Online]. Available:
  \url{http://www2.ece.ohio-state.edu/~schniter/pdf/basp17\_poster.pdf}
\BIBentrySTDinterwordspacing

\bibitem{Sung2013}
\BIBentryALTinterwordspacing
K.~Sung, B.~L. Daniel, and B.~A. Hargreaves, ``{Location Constrained
  Approximate Message Passing for Compressed Sensing MRI},'' \emph{Magnetic
  Resonance in Medicine}, vol.~70, pp. 370--381, 2013. [Online]. Available:
  \url{https://onlinelibrary.wiley.com/doi/pdf/10.1002/mrm.24468}
\BIBentrySTDinterwordspacing

\bibitem{Eksioglu2018}
\BIBentryALTinterwordspacing
E.~M. Eksioglu and A.~K. Tanc, ``{Denoising AMP for MRI Reconstruction:
  BM3D-AMP-MRI},'' \emph{SIAM Journal on Imaging Sciences}, vol.~11, no.~3, pp.
  2090--2109, jan 2018. [Online]. Available:
  \url{https://epubs.siam.org/doi/10.1137/18M1169655}
\BIBentrySTDinterwordspacing

\bibitem{Pipe1999}
\BIBentryALTinterwordspacing
J.~G. Pipe and P.~Menon, ``{Sampling density compensation in MRI: Rationale and
  an iterative numerical solution},'' \emph{Magnetic Resonance in Medicine},
  vol.~41, no.~1, pp. 179--186, jan 1999. [Online]. Available:
  \url{http://doi.wiley.com/10.1002/\%28SICI\%291522-2594\%28199901\%2941\%3A1\%3C179\%3A\%3AAID-MRM25\%3E3.0.CO\%3B2-V}
\BIBentrySTDinterwordspacing

\bibitem{Pruessmann2001}
\BIBentryALTinterwordspacing
K.~P. Pruessmann, M.~Weiger, P.~B{\"{o}}rnert, and P.~Boesiger, ``{Advances in
  sensitivity encoding with arbitrary k-space trajectories},'' \emph{Magnetic
  Resonance in Medicine}, vol.~46, no.~4, pp. 638--651, oct 2001. [Online].
  Available: \url{http://www.ncbi.nlm.nih.gov/pubmed/11590639}
\BIBentrySTDinterwordspacing

\bibitem{gamerman2006markov}
D.~Gamerman and H.~F. Lopes, \emph{{Markov chain Monte Carlo: stochastic
  simulation for Bayesian inference}}.\hskip 1em plus 0.5em minus 0.4em\relax
  CRC Press, 2006.

\bibitem{chatterjee2018sample}
S.~Chatterjee, P.~Diaconis, and Others, ``{The sample size required in
  importance sampling},'' \emph{The Annals of Applied Probability}, vol.~28,
  no.~2, pp. 1099--1135, 2018.

\bibitem{Mousavi2013}
\BIBentryALTinterwordspacing
A.~Mousavi, A.~Maleki, and R.~G. Baraniuk, ``{Parameterless Optimal Approximate
  Message Passing},'' oct 2013. [Online]. Available:
  \url{http://arxiv.org/abs/1311.0035}
\BIBentrySTDinterwordspacing

\bibitem{Bayati2013}
M.~Bayati, M.~A. Erdogdu, and A.~Montanari, ``{Estimating LASSO risk and noise
  level},'' \emph{Advances in Neural Information Processing Systems}, pp. 1--9,
  2013.

\bibitem{Vonesch2008}
C.~Vonesch and M.~Unser, ``{A fast thresholded landweber algorithm for
  wavelet-regularized multidimensional deconvolution},'' \emph{IEEE
  Transactions on Image Processing}, vol.~17, no.~4, pp. 539--549, 2008.

\bibitem{Bayram2010a}
I.~Bayram and I.~W. Selesnick, ``{A subband adaptive iterative
  shrinkage/thresholding algorithm},'' \emph{IEEE Transactions on Signal
  Processing}, vol.~58, no. 3 PART 1, pp. 1131--1143, 2010.

\bibitem{Donoho1995}
\BIBentryALTinterwordspacing
D.~L. Donoho and I.~M. Johnstone, ``{Adapting to Unknown Smoothness via Wavelet
  Shrinkage},'' \emph{Journal of the American Statistical Association},
  vol.~90, no. 432, p. 1200, dec 1995. [Online]. Available:
  \url{https://www.jstor.org/stable/2291512?origin=crossref}
\BIBentrySTDinterwordspacing

\bibitem{Khare2012}
\BIBentryALTinterwordspacing
K.~Khare, C.~J. Hardy, K.~F. King, P.~A. Turski, and L.~Marinelli,
  ``{Accelerated MR imaging using compressive sensing with no free
  parameters},'' \emph{Magnetic Resonance in Medicine}, vol.~68, no.~5, pp.
  1450--1457, nov 2012. [Online]. Available:
  \url{http://www.ncbi.nlm.nih.gov/pubmed/22266597}
\BIBentrySTDinterwordspacing

\bibitem{maleki2013asymptotic}
A.~Maleki, L.~Anitori, Z.~Yang, and R.~G. Baraniuk, ``{Asymptotic analysis of
  complex LASSO via complex approximate message passing (CAMP)},'' \emph{IEEE
  Transactions on Information Theory}, vol.~59, no.~7, pp. 4290--4308, 2013.

\bibitem{Beck2009}
\BIBentryALTinterwordspacing
A.~Beck and M.~Teboulle, ``{A Fast Iterative Shrinkage-Thresholding Algorithm
  for Linear Inverse Problems},'' \emph{SIAM Journal on Imaging Sciences},
  vol.~2, no.~1, pp. 183--202, jan 2009. [Online]. Available:
  \url{http://epubs.siam.org/doi/10.1137/080716542}
\BIBentrySTDinterwordspacing

\bibitem{Vasanawala2011}
S.~S. Vasanawala, M.~J. Murphy, M.~T. Alley, P.~Lai, K.~Keutzer, J.~M. Pauly,
  and M.~Lustig, ``{Practical parallel imaging compressed sensing MRI: Summary
  of two years of experience in accelerating body MRI of pediatric patients},''
  \emph{Proceedings - International Symposium on Biomedical Imaging}, pp.
  1039--1043, 2011.

\bibitem{Guerquin-Kern2011}
M.~Guerquin-Kern, M.~Haberlin, K.~P. Pruessmann, and M.~Unser, ``{A fast
  wavelet-based reconstruction method for magnetic resonance imaging},''
  \emph{IEEE Transactions on Medical Imaging}, vol.~30, no.~9, pp. 1649--1660,
  2011.

\bibitem{Makinen2020}
\BIBentryALTinterwordspacing
Y.~M{\"{a}}kinen, L.~Azzari, E.~S{\'{a}}nchez-Monge, M.~Maggioni, A.~Danielyan,
  K.~Dabov, A.~Foi, V.~Katkovnik, and K.~Egiazarian, ``{Image and video
  denoising by sparse 3D transform-domain collaborative filtering},'' 2020.
  [Online]. Available:
  \url{http://www.cs.tut.fi/~foi/GCF-BM3D/index.html\#ref\_results}
\BIBentrySTDinterwordspacing

\bibitem{Tsai2000}
C.~M. Tsai and D.~G. Nishimura, ``{Reduced aliasing artifacts using
  variable-density k-space sampling trajectories},'' \emph{Magnetic Resonance
  in Medicine}, vol.~43, no.~3, pp. 452--458, 2000.

\bibitem{Lazarus2019}
C.~Lazarus, P.~Weiss, N.~Chauffert, F.~Mauconduit, L.~{El Gueddari},
  C.~Destrieux, I.~Zemmoura, A.~Vignaud, and P.~Ciuciu, ``{SPARKLING:
  variable-density k-space filling curves for accelerated T2*-weighted MRI},''
  \emph{Magnetic Resonance in Medicine}, vol.~81, no.~6, pp. 3643--3661, 2019.

\bibitem{Baraniuk2010}
\BIBentryALTinterwordspacing
R.~G. Baraniuk, V.~Cevher, M.~F. Duarte, and C.~Hegde, ``{Model-Based
  Compressive Sensing},'' \emph{IEEE Transactions on Information Theory},
  vol.~56, no.~4, pp. 1982--2001, apr 2010. [Online]. Available:
  \url{http://ieeexplore.ieee.org/document/5437428/}
\BIBentrySTDinterwordspacing

\bibitem{Liang2009}
\BIBentryALTinterwordspacing
D.~Liang, B.~Liu, J.~Wang, and L.~Ying, ``{Accelerating SENSE using compressed
  sensing},'' \emph{Magnetic Resonance in Medicine}, vol.~62, no.~6, pp.
  1574--1584, dec 2009. [Online]. Available:
  \url{http://www.ncbi.nlm.nih.gov/pubmed/19785017}
\BIBentrySTDinterwordspacing

\bibitem{Ahmad2020}
R.~Ahmad, C.~A. Bouman, G.~T. Buzzard, S.~Chan, S.~Liu, E.~T. Reehorst, and
  P.~Schniter, ``{Plug-and-Play Methods for Magnetic Resonance Imaging: Using
  Denoisers for Image Recovery},'' \emph{IEEE Signal Processing Magazine},
  vol.~37, no.~1, pp. 105--116, jan 2020.

\bibitem{LISTA2010}
K.~Gregor and Y.~LeCun, ``{Learning Fast Approximations of Sparse Coding},'' in
  \emph{Proceedings of the 27th International Conference on International
  Conference on Machine Learning}, ser. ICML'10.\hskip 1em plus 0.5em minus
  0.4em\relax Madison, WI, USA: Omnipress, 2010, pp. 399--406.

\bibitem{NIPS2017_6774}
C.~Metzler, A.~Mousavi, and R.~Baraniuk, ``{Learned D-AMP: Principled Neural
  Network based Compressive Image Recovery},'' in \emph{Advances in Neural
  Information Processing Systems 30}, I.~Guyon, U.~V. Luxburg, S.~Bengio,
  H.~Wallach, R.~Fergus, S.~Vishwanathan, and R.~Garnett, Eds.\hskip 1em plus
  0.5em minus 0.4em\relax Curran Associates, Inc., pp. 1772--1783.

\bibitem{Ito2019}
D.~Ito, S.~Takabe, and T.~Wadayama, ``{Trainable ISTA for sparse signal
  recovery},'' \emph{IEEE Transactions on Signal Processing}, vol.~67, no.~12,
  pp. 3113--3125, 2019.

\bibitem{Metzler2018}
\BIBentryALTinterwordspacing
C.~A. Metzler, A.~Mousavi, R.~Heckel, and R.~G. Baraniuk, ``{Unsupervised
  Learning with Stein's Unbiased Risk Estimator},'' 2018. [Online]. Available:
  \url{http://arxiv.org/abs/1805.10531}
\BIBentrySTDinterwordspacing

\bibitem{Zhussip2019}
M.~Zhussip, S.~Soltanayev, and S.~Y. Chun, ``{Training deep learning based
  image denoisers from undersampled measurements without ground truth and
  without image prior},'' \emph{Proceedings of the IEEE Computer Society
  Conference on Computer Vision and Pattern Recognition}, vol. 2019-June, pp.
  10\,247--10\,256, 2019.

\bibitem{Luisier2007}
F.~Luisier, T.~Blu, and M.~Unser, ``{A new SURE approach to image denoising:
  Interscale orthonormal wavelet thresholding},'' \emph{IEEE Transactions on
  Image Processing}, vol.~16, no.~3, pp. 593--606, 2007.

\end{thebibliography}

\end{document}